%% file: main_arXiv.tex
\documentclass[pdflatex,sn-mathphys-num]{sn-jnl}

\usepackage{graphicx}%
\usepackage{multirow}%
\usepackage{amsmath,amssymb,amsfonts}%
\usepackage{amsthm}%
\usepackage{mathrsfs}%
\usepackage[title]{appendix}%
\usepackage{xcolor}%
\usepackage{textcomp}%
\usepackage{manyfoot}%
\usepackage{booktabs}%
\usepackage{algorithm}%
\usepackage{algorithmicx}%
\usepackage{algpseudocode}%
\usepackage{listings}%

\usepackage{hyperref}

\usepackage[inkscapelatex=false]{svg}

\usepackage{color,soul,xcolor}
\definecolor{darkgreen}{rgb}{0.0, 0.5, 0.0} 

\usepackage{multirow}
\usepackage{caption}
\usepackage{subcaption}

\usepackage{amsmath,amsthm,amssymb,amsfonts}

\usepackage{algorithm}
\usepackage{algpseudocode}

\usepackage{listings}
\usepackage{xcolor}

\usepackage{minted}
\definecolor{vscodeBackground}{RGB}{245,245,245} 
\definecolor{vscodeKeyword}{RGB}{0,128,0}        
\definecolor{vscodeFunction}{RGB}{0,0,180}       
\definecolor{vscodeType}{RGB}{43,145,175}        
\definecolor{vscodeComment}{RGB}{0,128,0}        
\definecolor{vscodeString}{RGB}{163,21,21}       
\definecolor{vscodeNumber}{RGB}{128,0,128}       

\setminted{
    linenos,
    breaklines=true,
    breaksymbol={},                  
    frame=lines,
    bgcolor=vscodeBackground,
    fontsize=\small,
    tabsize=4,
    escapeinside=||,               
}

\usepackage[utf8]{inputenc}
\DeclareUnicodeCharacter{2606}{\ensuremath{\star}}
\DeclareUnicodeCharacter{03B1}{\ensuremath{\alpha}}

\theoremstyle{thmstyleone}%
\theoremstyle{thmstyletwo}%
\newtheorem{remark}{Remark}[section]%

\theoremstyle{thmstylethree}%

\begin{document}

\title[An open-source computational framework for immersed fluid-structure interaction modeling using FEBio and MFEM]{An open-source computational framework for immersed fluid-structure interaction modeling using FEBio and MFEM}


\author*[1,2]{\fnm{Ryan T.} \sur{Black}}\email{blackr1@chop.edu}
\equalcont{These authors contributed equally to this work.}
\author[3,4]{\fnm{Steve A.} \sur{Maas}}
\equalcont{These authors contributed equally to this work.}
\author[1,5,6]{\fnm{Wensi} \sur{Wu}}
\author[1]{\fnm{Jalaj} \sur{Maheshwari}}
\author[7]{\fnm{Tzanio} \sur{Kolev}}
\author*[3,4]{\fnm{Jeffrey A.} \sur{Weiss}}\email{jeff.weiss@utah.edu}
\author*[1,2,6]{\fnm{Matthew A.} \sur{Jolley}}\email{jolleym@chop.edu}

\affil[1]{\orgname{Children's Hospital of Philadelphia}, \orgdiv{Department of Anesthesiology and Critical Care Medicine}, \orgaddress{\city{Philadelphia}, \state{PA}, \country{USA}}}

\affil[2]{\orgname{Children's Hospital of Philadelphia}, \orgdiv{Division of Cardiology}, \orgaddress{\city{Philadelphia}, \state{PA}, \country{USA}}}

\affil[3]{\orgname{University of Utah}, \orgdiv{Department of Biomedical Engineering}, \orgaddress{\city{Salt Lake City}, \state{UT}, \country{USA}}}

\affil[4]{\orgname{University of Utah}, \orgdiv{Scientific Computing Institute}, \orgaddress{\city{Salt Lake City}, \state{UT}, \country{USA}}}

\affil[5]{\orgname{University of Pennsylvania}, \orgdiv{Department of Mechanical Engineering and Applied Mechanics}, \orgaddress{\city{Philadelphia}, \state{PA}, \country{USA}}}

\affil[6]{\orgname{Children's Hospital of Philadelphia}, \orgdiv{Cardiovascular Institute}, \orgaddress{\city{Philadelphia}, \state{PA}, \country{USA}}}

\affil[7]{\orgname{Lawrence Livermore National Laboratory}, \orgdiv{Center for Applied Scientific Computing}, \orgaddress{\city{Livermore}, \state{CA}, \country{USA}}}


\abstract{Fluid-structure interaction (FSI) simulation of biological systems presents significant computational challenges, particularly for applications involving large structural deformations and contact mechanics, such as heart valve dynamics. Traditional arbitrary Lagrangian-Eulerian methods encounter fundamental difficulties with such problems due to mesh distortion, motivating immersed techniques. This work presents a novel open-source immersed FSI framework that strategically couples two mature finite element libraries: MFEM, a GPU-ready and scalable library with state-of-the-art parallel performance developed at Lawrence Livermore National Laboratory, and FEBio, a nonlinear finite element solver with sophisticated solid mechanics capabilities designed for biomechanics applications developed at the University of Utah and Columbia University. This coupling creates a unique synergy wherein the fluid solver leverages MFEM's distributed-memory parallelization and pathway to GPU acceleration, while the immersed solid exploits FEBio's comprehensive suite of hyperelastic and viscoelastic constitutive models and advanced solid mechanics modeling targeted for biomechanics applications. FSI coupling is achieved using a fictitious domain/distributed Lagrange multiplier methodology with variational multiscale stabilization for enhanced accuracy on under-resolved grids expected with unfitted meshes used in immersed FSI. A fully implicit, monolithic scheme provides robust coupling for strongly coupled fluid-solid interactions characteristic of cardiovascular applications. The framework's modular architecture facilitates straightforward extension to additional physics and element technologies. Several test problems are considered to demonstrate the capabilities of the proposed framework, including a three-dimensional semilunar heart valve simulation. This platform addresses a critical need for open-source immersed FSI software combining advanced biomechanics modeling with high-performance computing infrastructure.}

\keywords{Fluid-structure interaction, Immersed method, Heart valve, MFEM, FEBio, Open-source software}

\maketitle

\section{Introduction}\label{sec:intro}

Fluid–structure interaction (FSI) encompasses multi-physics phenomena that arise when deforming or moving solids interact with surrounding fluids, a scenario that pervades biomechanics. Applications span blood flow in compliant vessels \cite{baumler_fluidstructure_2020,hirschhorn_fluid_2020,hu_multiphysics_2025,rostam-alilou_fluidstructure_2022,chen_fluidstructure_2021,wang_fluid-structure_2022}, heart valve dynamics \cite{zakerzadeh_computational_2017,sotiropoulos_review_2009,sun_computational_2014,lee_fluidstructure_2020,kaiser_simulation-based_2024,choi_computational_2014,johnson_effects_2022,tezduyar_immersogeometric_2018,nestola_fully_2021}, airway and respiratory mechanics \cite{luo_immersed-boundary_2008,malve_fsi_2011,malve_fsi_2011-1,chen_computational_2023,li_fluid-structure_2023,pirnar_computational_2015}, cell–fluid interactions \cite{azizi_simulating_2023,rauch_coupled_2018,scheiner_poromicromechanics_2016}, joint lubrication \cite{noori-dokht_finite_2017,moghani_finite_2007}, and various medical devices \cite{bazilevs_patient-specific_2009,fraser_use_2011,gay_stent_2006,xu_framework_2018,long_shape_2014,fiore_development_2002}.  
Biomechanical FSI presents significant challenges for computational modeling due to strongly coupled and highly nonlinear behavior, large solid deformations, complex and often anisotropic constitutive models, the presence of thin flexible structures, and contact or topological changes in the fluid domain. These factors make accurate and robust simulation of biomechanical FSI particularly demanding for numerical solvers.

FSI simulation of the cardiovascular system presents computational challenges arising from the coupling between pulsatile blood flow and compliant vascular tissues, requiring simultaneous resolution of large structural deformations, complex geometries, rich flow physics, and nonlinear material behavior \cite{taylor_patient-specific_2009,hirschhorn_fluid_2020}. These challenges pervade cardiovascular biomechanics—from arterial wall mechanics and aneurysm rupture prediction \cite{humphrey_mechanics_1995,humphrey_mechanics_2012} to ventricular assist device optimization \cite{bazilevs_patient-specific_2009,fraser_use_2011} and vascular graft design \cite{sankaran_nanoarchitecture_2015,szafron_optimization_2019}. Cardiac valves, however, exemplify these challenges in their most demanding form: leaflets undergo extreme deformations and experience contact during coaptation, which results in a topological change of the fluid domain, all within milliseconds \cite{sacks_heart_2007,sacks_biomechanics_2009,sacks_simulation_2019}. Accurate valve simulation must resolve both the hemodynamic consequences of valve function (regurgitation, pressure gradients, flow patterns) and the tissue-level stress and strain distributions that govern long-term durability and pathophysiological remodeling—a dual requirement that places valve FSI at the frontier of computational biomechanics \cite{ayoub_heart_2016,arminio_fluid-structure_2024}.

Simulations of pediatric populations amplify these demands considerably. Congenital heart disease affects approximately 1 percent of live births and frequently involves valve malformations \cite{hoffman_incidence_2002}. In children, native valve repair is strongly preferred over replacement because prosthetic valves cannot accommodate somatic growth, inevitably leading to patient-prosthesis mismatch and repeated reoperations, each carrying cumulative mortality risk \cite{alsoufi_aortic_2009,henaine_valve_2012}. Mechanical valves require lifelong anticoagulation that is difficult to manage in active children, while bioprosthetic valves demonstrate accelerated structural degeneration in younger patients with durability inversely related to age \cite{alsoufi_aortic_2009}. These constraints place a premium on optimizing native valve repair techniques. The heterogeneous anatomies arising from diverse congenital morphologies, combined with small patient populations that preclude empirical optimization through large clinical cohorts, motivate patient-specific computational approaches for individualized repair planning \cite{wu_computational_2022}.

When repair is not feasible, improved prosthetic valve designs remain essential for both pediatric and adult populations. Simulation of bioprosthetic valve mechanics is critical for predicting fatigue life, calcification propensity, and long-term durability \cite{martin_comparison_2015,kostyunin_degeneration_2020}. Similarly, the rapidly expanding use of transcatheter valve replacement demands computational tools to evaluate device performance across diverse patient anatomies. In these contexts, accurate tissue-level stress analysis—not merely hemodynamic assessment—is necessary to guide device optimization and predict failure modes \cite{yoganathan_flow_2005}.

Contemporary computational FSI methods can be broadly classified into boundary-fitted and non-boundary fitted approaches. In boundary-fitted techniques, the fluid mechanics problem is posed on a moving domain, utilizing the arbitrary Lagrangian Eulerian (ALE) description, to allow the fluid mesh to align with the solid mesh at the fluid–solid interface, which results in very accurate modeling of FSI problems. ALE-FSI methods, while capable of high boundary accuracy, encounter fundamental difficulties with problems involving large deformations and topological changes in the fluid domain (e.g. contact), due to the possibility of significant mesh distortion requiring remeshing, which can be expensive and introduce additional errors \cite{johnson_mesh_1994}. These limitations can make accurate modeling of the complex interaction between pulsatile blood flow and thin leaflets that undergo contact and large deformations throughout the cardiac cycle challenging and computationally expensive \cite{bavo_fluid-structure_2016,kheradvar_emerging_2015, lee_fluidstructure_2020}. Similar limitations arise in other cardiovascular and biomechanics applications involving large motions or evolving geometries, including ventricular assist device impellers with rotating components \cite{fraser_use_2011}, collapsible vessels \cite{tang_simulating_2002}, and stent–vessel interactions \cite{gay_stent_2006}.

Such difficulties have motivated non-boundary fitted FSI approaches, including the immersed boundary method \cite{peskin_flow_1972,peskin_immersed_2002,griffith_adaptive_2007,griffith_hybrid_2017,boffi_hyper-elastic_2008,mittal_versatile_2008,seo_sharp-interface_2011,borazjani_fluidstructure_2013}, immersed finite element method \cite{zhang_immersed_2004,wang_semi-implicit_2012,wang_modified_2013}, fictitous domain/distributed Lagrange multiplier method \cite{black_immersed_2025,boffi_finite_2015,boffi_distributed_2018,hesch_continuum_2012,hesch_mortar_2014,casquero_nurbs-based_2015,yu_dlmfd_2005,glowinski_distributed_1999,baaijens_fictitious_2001,de_hart_two-dimensional_2000,boilevin-kayl_loosely_2019}, cut finite element methods and extended finite element method (XFEM) based techniques \cite{gerstenberger_extended_2008,schott_monolithic_2019,boilevin-kayl_numerical_2019}, the shifted-boundary method \cite{xu_weighted_2024,xu_weighted_2026}, meshfree methods such as smoothed particle hydrodynamics \cite{mao_fully-coupled_2017,laha_smoothed_2024,fourey_efficient_2017}, and immersogeometric analysis \cite{kamensky_immersogeometric_2015}, which naturally accommodate complex solid motions and deformations without a moving fluid mesh. This class of computational FSI techniques immerse or embed the solid within a fixed background fluid domain, thereby avoiding the need for specialized procedures to maintain mesh quality as required in boundary-fitted formulations. However, since the fluid grid is neither aligned with the fluid–structure interface nor updated to track solid motion, the standard immersed FSI approach typically exhibits reduced accuracy in the vicinity of the interface compared with ALE-FSI techniques. This drawback can be mitigated through adaptive mesh-refinement strategies that locally enhance resolution near the interface \cite{griffith_adaptive_2007}, as well as discretization approaches, such as the variational multiscale (VMS) method \cite{bazilevs_variational_2007,bazilevs_isogeometric_2012}, that improve accuracy and robustness on under-resolved grids, an expected challenge with unfitted meshes employed in immersed FSI techniques \cite{black_immersed_2025}. 

In this work, our motivating application is computational FSI analysis of pediatric heart valve dynamics. A variety of FSI techniques have been applied to simulate the interaction between pulsatile blood flow and heart valves. Although ALE‑FSI techniques provide high accuracy near the fluid–solid interface, these methods have realized limited success in simulating heart valve FSI dynamics due to the difficulty of modeling the large motions of thin leaflet structures with conforming meshes, which necessitates frequent remeshing and sophisticated contact modeling, ultimately increasing computational cost \cite{lee_fluidstructure_2020, kheradvar_emerging_2015}. Bavo et al. compared an immersed boundary method with an ALE-FSI approach for simulating valve dynamics and reported that the ALE-FSI method was unable to successfully simulate the full cardiac cycle because of remeshing difficulties \cite{bavo_fluid-structure_2016}. In contrast, immersed methods that avoid the difficulties with body-fitted grids have been widely used to model heart valve dynamics, including problems such as bioprosthetic heart valve design \cite{hsu_dynamic_2015,lee_fluidstructure_2020,nestola_fully_2021,borazjani_fluidstructure_2013,xu_framework_2018,oks_fluidstructure_2022,choi_computational_2014}, analysis and optimization of surgical interventions \cite{kaiser_simulation-based_2024,choi_effect_2024,wu_immersogeometric_2019}, and investigation of native, repaired, and diseased leaflet mechanics and hemodynamics \cite{ko_computational_2024,fringand_analysis_2024,seo_flow_2020,kaiser_comparison_2023,bornemann_leaflet_2025,abbas_closure_2025}. 

To date, the majority of open-source biomechanics and cardiovascular simulation software packages rely on ALE or body-fitted formulations for FSI modeling. SimVascular \cite{updegrove_simvascular_2017} and CRIMSON \cite{arthurs_crimson_2021} support FSI simulations of vessel‑wall deformation using the coupled momentum method, a reduced FSI technique that treats the vessel wall as a linear membrane enhanced with transverse shear \cite{figueroa_coupled_2006}. SimVascular additionally offers an ALE approach for FSI simulations. FEBio provides an ALE‑FSI capability based on a novel fluid mechanics framework that uses dilatation as the primary variable instead of pressure \cite{ateshian_finite_2018, shim_formulation_2019}. The life$^x$-cfd framework, built on the deal.II finite element library \cite{arndt_dealii_2025}, includes support for one‑way FSI via prescribed solid motion using an ALE description of the moving fluid domain, as well as an immersed formulation, the resistive immersed implicit surface (RIIS) method, for modeling moving rigid structures embedded in the fluid domain \cite{africa_lifex-cfd_2024}.

Current open-source immersed FSI solvers generally offer more limited modeling capabilities than the aforementioned simulation software packages. IBAMR is an adaptive, distributed‑memory parallel implementation of the immersed boundary method that has been widely used in biomechanics applications, including heart valve FSI analysis \cite{griffith_ibamr_nodate,lee_fluidstructure_2020,davey_simulating_2024,kaiser_design-based_2021,kaiser_comparison_2023}. General purpose open‑source solvers and libraries, including OpenFOAM \cite{wu_numerical_2014} and FEniCS \cite{kamensky_open-source_2021,neighbor_leveraging_2023,hirschvogel_ambit_2024,bourantas_immersed_2021}, have also been applied to biological FSI problems. These platforms, though flexible, often require substantial user implementation effort to incorporate advanced material laws, contact mechanics, or fully coupled FSI workflows necessary to simulate heart valve dynamics. Nevertheless, ParaValve \cite{saraeian_paravalve_2025}, which is built on the FEniCS based immersed FSI solver from \cite{neighbor_leveraging_2023}, provides a complete pipeline to simulate parametric bioprosthetic valves within a patient-specific aorta geometry.

In summary, most existing open-source immersed FSI solvers lack the sophisticated solid mechanics modeling features and specialized fluid boundary conditions necessary for modeling valve dynamics under physiological conditions. Furthermore, user extensions of these open-source solvers that utilize reduced-order or rigid solid models can be effective for investigating the hemodynamic parameters of valve performance, however these simplified structural representations inadequately capture the tissue mechanics fundamental to understanding disease progression, calcific degeneration, mechanobiological response, and device-tissue interactions \cite{ayoub_heart_2016,el-tallawi_valve_2021,el-tallawi_mitral_2021,marsan_valve_2021,narang_pre-surgical_2021,meador_tricuspid_2020}. Additionally, frameworks built on top of general purpose solver libraries are typically specialized for one application problem and would require further user development to be applied to a wider range of valve FSI problems. Beyond methodological and modeling considerations, the computational cost of high-fidelity FSI simulations has historically limited their practical utility; however, emerging computing architectures leveraging hardware accelerators, e.g. graphics processing units (GPUs), are increasingly transforming this landscape, yielding significant speed-up compared to traditional CPU-based approaches \cite{andrej_high-performance_2024,kumar_gpu-accelerated_2025}. As such, there remains a critical need for an open-source immersed FSI framework that combines support for large-deformation cardiovascular problems with sophisticated solid modeling capabilities, including hyperelastic constitutive models, contact algorithms, and specialized element formulations, that can leverage modern parallel computing architectures for practical simulation times.

To address this need, we present a novel open-source immersed FSI framework coupling the modular finite element methods (MFEM) library \cite{anderson_mfem_2021}, a scalable finite element library supporting massive parallelization via distributed memory computing and hardware accelerators (e.g. GPUs), with FEBio \cite{maas_febio_2012}, a nonlinear finite element solver designed specifically for biomechanical applications. FEBio provides sophisticated solid mechanics capabilities including hyperelastic and viscoelastic constitutive models appropriate for soft tissues, robust contact algorithms essential for leaflet coaptation and tissue-device interaction, specialized boundary conditions for cardiovascular analysis, as well as capabilities for modeling transport, chemical reactions, growth, and remodeling \cite{maas_febio_2012,ateshian_finite_2018,shim_formulation_2019,ateshian_continuum_2024,shim_finite_2023,ateshian_computational_2014,wu_computational_2022}. This immersed FSI framework overcomes the inherent limitations of ALE‑FSI methods for valve applications, enabling the study of pulsatile hemodynamics alongside the complex stress and strain fields in valve tissues necessary to understanding disease progression and mechanobiological tissue response. Although valve FSI modeling serves as the motivating application, the modular architecture of this framework accommodates diverse biomechanical and cardiovascular FSI applications. By providing an open-source platform with extensible workflows, this work aims to fill a critical gap in the computational biomechanics community, enabling integrated studies that address both fluid dynamic performance and tissue-level mechanical response across a broad spectrum of biomedical applications, including those with large solid deformations and motions.

The paper is organized as follows. In Section \ref{sec:methods}, we describe the mathematical formulation and numerical algorithms of the immersed FSI technique used in our solver. Implementation details and design of the computational framework coupling two C++ finite element libraries are described in Section \ref{sec:plugin_design}. We present several numerical examples in Section \ref{sec:tests} to verify the implementation, as well as demonstrate the potential of the computational FSI tool to simulate a wide range of FSI problems, including three-dimensional heart valve dynamics. Lastly, a discussion of the open-source framework is provided in Section \ref{sec:discussion} and concluding remarks are given in Section \ref{sec:conclusion}.

\section{Immersed FSI Method}\label{sec:methods}
This section presents the immersed FSI method used by the framework, which is based on the approach presented in \cite{black_immersed_2025} with some modifications described below to facilitate simpler coupling between FEBio and MFEM as well as improve accuracy when modeling large pressure jumps across the FSI interface. We refer the reader to the prior work and citations therein for a more detailed derivation and presentation of the method, which we omit here for the sake of brevity.

In this work, we consider a solid fully or partially immersed in a background fluid domain. Let $\Omega = \Omega^f_t \cup \Omega^s_t$ denote the domain of interest composed of fluid and solid domains respectively, where $t\in[0,T]$ signifies the configuration of the body at time $t$. In the current implementation, we focus on a fixed domain $\Omega$, however the present work could be generalized to a moving domain leveraging the arbitrary Langrangian-Eulerian (ALE) technique resulting in a combined ALE and immersed FSI approach. Let the boundaries of each subdomain be denoted as $\Gamma^{(\cdot)}_{t,i}$ where $i=D,N$ for Dirichlet and Neumann boundaries and let $\Gamma^{fsi}_t$ signify the fluid-solid interface with outward normals $\mathbf{n}^f$ and $\mathbf{n}^s$.

The present immersed FSI approach falls into the family of fictitious domain (FD) methods (see e.g. \cite{glowinski_distributed_1999,yu_dlmfd_2005,boffi_finite_2015}). In this class of techniques, the fluid equations are extended to the entire domain $\Omega$ and the solid is immersed or embedded into the background fluid domain. This approach introduces artificial fluid within the overlapping region, which can be accounted for by subtracting the work done by this artificial fluid from the solid equations. The coupling between the fluid and immersed solid is achieved in the following way: (1) the continuity of tractions along the FSI interface is naturally satisfied by the variational equations, and (2) the condition of matching fluid and solid velocity along $\Gamma^{fsi}$ has been transformed, via the FD method, to a condition on the overlapping region using a distributed Lagrange multiplier. It is the latter that removes the need for an explicit interface mesh that moves/deforms with the fluid and solid domains, an attractive feature for modeling large deformation FSI problems. The immersed FSI variational problem is formulated as follows: Find $(\mathbf{v}^f, \dot{\mathbf{u}}^s, p, \boldsymbol{\lambda})\in\mathcal{S}_f\times\mathcal{S}_s\times\mathcal{S}_p\times\mathcal{S}_{\lambda}$ such that $\forall(\mathbf{w}^f, \mathbf{w}^s, q, \delta\boldsymbol{\lambda})\in\mathcal{V}_f\times\mathcal{V}_s\times\mathcal{V}_p\times\mathcal{V}_{\lambda}$,
\begin{align}
    & R_{f}(\{\mathbf{w}^f,q\},\{\mathbf{v}^f,p\}) + R_s(\mathbf{w}^s,\dot{\mathbf{u}}^s) \nonumber \\
    & + C_{\lambda}(\mathbf{w}^s,\mathbf{w}^f,\boldsymbol{\lambda}) + C_{\lambda}(\dot{\mathbf{u}}^s,\mathbf{v}^f,\delta\boldsymbol{\lambda}) = 0,
    \label{eq:fsi_weak_form}
\end{align}
where $(\mathbf{v}^f, \dot{\mathbf{u}}^s, p, \boldsymbol{\lambda})$ are the fluid velocity, solid velocity, fluid pressure, and Lagrange multiplier to enforce the volumetric constraint; $(\mathbf{w}^f, \mathbf{w}^s, q, \delta\boldsymbol{\lambda})$ are the corresponding test functions; $\mathcal{S}_{(\cdot)}$ are the trial solution spaces that satisfy the Dirichlet boundary conditions; and $\mathcal{V}_{(\cdot)}$ are the test function spaces that satisfy homogenous Dirichlet boundary conditions where the trial solutions have Dirichlet boundary conditions.

\subsection{Fluid}
In this work, we model the fluid as incompressible and Newtonian. The weak form for the conservation of mass and momentum (i.e. the incompressible Navier-Stokes equations) are given by
\begin{align}
    & R_{f}(\{\mathbf{w}^f,q\},\{\mathbf{v}^f,p\}) = \int_{\Omega} \mathbf{w}^f \cdot \rho^f\left( \dot{\mathbf{v}}^f - \mathbf{f} \right) \: d\Omega - \int_{\Omega} p \: \nabla\cdot \mathbf{w}^f \; d\Omega \nonumber \\
    & + \int_{\Omega} 2\mu^f \: \boldsymbol{\epsilon}(\mathbf{w}^f) : \:\boldsymbol{\epsilon}(\mathbf{v}^f) \; d\Omega - \int_{\Gamma^{f}_{t,N}} \mathbf{w}^f \cdot \mathbf{h}_f \; d\Gamma \nonumber \\
    & + \int_{\Omega}q\nabla\cdot\mathbf{v}^f\: d\Omega,
    \label{eq:fluid_weak_form}
\end{align}
where $\rho^f$ is the fluid density, $\mu^f$ is the fluid dynamic viscosity, $\mathbf{f}$ is a given body force, and $\mathbf{h}_f$ is a given surface traction on $\Gamma^f_{t,N}$. In the present implementation, we focus on Newtonian fluids, however future work will extend the framework to non-Newtonian fluids. 

We discretize the above weak form using finite dimensional subspaces of the trial and test function spaces, and augment the equations with additional terms following the Variational Multiscale (VMS) method \cite{bazilevs_variational_2007,bazilevs_isogeometric_2008}. We omit the VMS weak form for brevity and refer the reader to \cite{black_immersed_2025} for details. These stabilization terms and coefficients provide enhanced stability and accuracy on under-resolved grids (typical for unfitted FSI and advection-dominated flows), ensure optimal convergence rates with respect to mesh size and polynomial order, and allow for the use of computationally efficient equal-order interpolants for the velocity and pressure trial solution spaces \cite{bazilevs_computational_2013}. Furthermore, we utilize a modification of the stabilization coefficients originally presented in \cite{kamensky_immersogeometric_2015} and leveraged in subsequent works \cite{casquero_nurbs-based_2017,boilevin-kayl_numerical_2019} as follows
\begin{align}
    & \tau_m = \left( s \left( \frac{C_T}{\Delta t^2} + \mathbf{v}^f\cdot\mathbf{G}\:\mathbf{v}^f + C_I\left(\frac{\mu^f}{\rho} \right)^2\mathbf{G}:\:\mathbf{G} \right) \right)^{-1/2}, \\
    & \tau_c = (\tau_m \: tr\mathbf{G})^{-1}, \\
    & G_{ij} = \sum_{k=1}^{n_{sd}} \frac{\partial\xi_k}{\partial x_i} \frac{\partial\xi_k}{\partial x_j},
\end{align}
where $n_{sd}$ is the number of spatial dimensions, $C_I$ is a constant related to the order of polynomials used in the finite element approximation, taken to be $36$ for linear basis functions which is derived from element-wise inverse estimates \cite{franca_stabilized_1992,johnson_numerical_2009,bazilevs_variational_2007}, $C_T$ is a constant specific to the time-stepping method, in this case set to $4$ following \cite{bazilevs_variational_2007,bazilevs_isogeometric_2008}, and $\boldsymbol{\xi}$ are the coordinates of the parametric domain of an element. The additional scaling factor, $s$, is introduced to enhance the numerical approximation in the vicinity of the fluid–solid interface, where the pressure gradient is poorly approximated due to the use of a continuous pressure approximation, potentially resulting in poor mass conservation. This will be especially problematic for the case of immersed thin structures, which are known to have large pressure jumps across the solid. The value of $s$ is allowed to vary in space and typically chosen to have a value $s>>1$ in an $O(h)$ neighborhood around the fluid-solid interface and $s=1$ otherwise, with a smooth transition between the regions achieved by defining $s$ at the fluid nodes and interpolating with the same space as the pressure finite element approximation. This results in a local reduction of the value of $\tau_m$ around the interface and thus the effect of a poor pressure gradient approximation is diminished, while simultaneously boosting volume loss penalization due to the reciprocal relationship between $\tau_m$ and $\tau_c$.

\subsection{Solid}
We primarily focus on hyperelastic solids immersed in fluid, however the formulation is not limited to this case as no assumptions were made about the constitutive behavior of the solid when deriving the immersed FSI formulation as described in \cite{black_immersed_2025}. Recall, for the present FD type FSI approach, there exists artificial fluid within the overlapping region, which we account for by subtracting the work done by this artificial fluid from the solid equations. The resulting weak form for the immersed solid is as follows,
\begin{align}
    & R_{s}(\mathbf{w}^s,\dot{\mathbf{u}}^s) = \int_{\Omega^s_t} \mathbf{w}^s \cdot (\rho^s - \rho^f)\left(\ddot{\mathbf{u}}^s - \mathbf{f}\right)\: d\Omega + \int_{\Omega^s_t}\nabla\mathbf{w}^s:\boldsymbol{\sigma}^s\:d\Omega \nonumber \\
    &  - \int_{\Omega^s_t} 2\mu^f\boldsymbol{\epsilon}(\mathbf{w}^s):\boldsymbol{\epsilon}(\mathbf{v}^f)\: d\Omega - \int_{\Gamma^s_{t,N}} \mathbf{w}^s\cdot\mathbf{h}_s\:d\Gamma,
    \label{eq:solid_weak_form}
\end{align}
where $\rho^s$ is the solid density, $\mathbf{f}$ is a given body force, $\mathbf{h}_s$ is a given surface traction on $\Gamma^s_{t,N}$, $\boldsymbol{\sigma}^s$ is the solid Cauchy stress, and $\boldsymbol{\epsilon}(\cdot) = 0.5\left( \nabla(\cdot) + \nabla(\cdot)^T \right)$. The Cauchy stress is defined by a strain-energy function per unit volume $\psi$,
\begin{equation}
    \boldsymbol{\sigma}^s = \frac{2}{J}\mathbf{F}\frac{\partial\psi(\mathbf{C})}{\partial\mathbf{C}}\mathbf{F}^T,
\end{equation}
where $\mathbf{F}= \frac{\partial \mathbf{x}}{\partial \mathbf{X}}$, $\mathbf{x}$ denotes a point in the current configuration, $\mathbf{X}$ denotes a point in the reference configuration, $J = \text{det}\mathbf{F}$, and $\mathbf{C} = \mathbf{F}^T\mathbf{F}$ is the right Cauchy-Green deformation tensor.

Note the presence of the density difference between the solid and fluid in the first term, as well as the subtraction of the fluid viscous stress in the last term, which represents the work done by the artificial fluid in the overlapping region. Here, we do not subtract the full fluid stress tensor since we make the additional assumption that the immersed solid is incompressible or at least nearly-incompressible. This simplifies the coupling between the fluid and solid as both are modeled as incompressible or nearly-incompressible. As a result, we can leverage the combination of the incompressibility constraint in the fluid that is satisfied in the entire domain $\Omega$ along with the condition of matching fluid and solid velocity in the overlapping region to enforce incompressibility in the solid. In principle, for incompressible solids, only the isochoric deformations need consideration as $J=1$. However, errors in approximating the velocity fields, interpolation errors between the two unfitted meshes, and weak enforcement of the incompressibility constraint in the fluid can produce artificial solid stresses due to incompressibility errors. In general, we keep the volumetric response in the solid constitutive model with the interpretation that these additional terms produce stresses that counteract violations of incompressibility. For more details on this, see \cite{black_immersed_2025}.

\begin{remark}
    It is possible with the present immersed FSI formulation to model compressible solids immersed in incompressible fluid by modifying the incompressibility constraint within the overlapping region to allow for non-isochoric deformations (e.g. \cite{liu_immersed_2006,hesch_continuum_2012,heltai_variational_2012,boffi_distributed_2018,wang_modified_2013}).
\end{remark}

Upon selection of finite dimensional subspaces of the trial and test function spaces, the above weak form provides a starting point for our immersed solid spatial discretization. Due to the coupling with FEBio, the solver has the potential to access a wide range of constitutive models and element technologies \cite{maas_febio_2012}. For example, the plugin can interface with the existing FEBio implementation of the mean dilation method for modeling nearly-incompressible materials typically used for biological soft-tissues \cite{simo_quasi-incompressible_1991}. See \cite{maas_febio_2012,noauthor_febio_nodate} and references therein for a more detailed description of the capabilities of the FEBio software.

\subsection{FSI Constraint} \label{subsec:fsi_constraint}
The volumetric constraint of matching fluid and solid velocity in the overlapping region is given by
\begin{equation}
    C_{\lambda}(\dot{\mathbf{u}}^s,\mathbf{v}^f,\delta\boldsymbol{\lambda}) = \int_{\Omega_t^s} \delta\boldsymbol{\lambda}\cdot\left(\dot{\mathbf{u}}^s - \mathbf{v}^f\right)\: d\Omega = 0,
    \label{eq:fsi_constraint}
\end{equation}
which can be obtained from Equation \ref{eq:fsi_weak_form} by setting $\mathbf{w}^s = \mathbf{w}^f = \mathbf{0}$. The above involves the integration over the immersed solid domain of the difference between $\dot{\mathbf{u}}^s\in\mathcal{S}_s$ defined on the solid domain and $\mathbf{v}^f\in\mathcal{S}_f$ defined on the fluid domain. With a special choice for the Lagrange multiplier finite element space, namely letting the basis functions be delta functions at the solid nodes ($N^{\lambda}_A = \delta(\mathbf{X}-\mathbf{X}_A)$), the above integral can be simplied to 
\begin{equation}
    \dot{\mathbf{u}}^s_A = \sum_{B\in\omega_v}N^v_B(\boldsymbol{\phi}(\mathbf{X}_A,t))\mathbf{v}^f_B,
    \label{eq:nodal_solid_velo}
\end{equation}
where we have substituted in finite element approximations of the form $\mathbf{x} = \sum_{A\in\omega_{x}} N^{x}_A \mathbf{x}_A$, where $(\cdot)_A$ denotes a quantity at node $A$ and $N^{(\cdot)}_A(\cdot)$ is the corresponding shape function, $\omega_{(\cdot)}$ is the set of nodes for a quantity excluding any nodes on Dirichlet boundaries, and $\boldsymbol{\phi}(\bold{X},t)$ is the motion map from the reference to current configuration. This constraint is now strongly enforced at the solid nodes, and consequently only one velocity field exists in our formulation. The solid velocity is simply a projection, via nodal interpolation, of the fluid velocity to the immersed solid domain. Although this may seem counterintuitive to match the velocities between two distinct material behaviors, recall the fluid in the overlapping solid region is artificial and we constrain this artificial fluid behavior to match that of the immersed solid body.

\begin{remark}
    As described in \cite{hesch_mortar_2014}, this choice of Lagrange multiplier finite element basis functions makes the present immersed FSI formulation equivalent to the immersed finite element method (IFEM) \cite{zhang_immersed_2004}.
\end{remark}

This choice of shape function for the Lagrange multiplier finite element space has further implications for the term 
\begin{equation}
    C_{\lambda}(\mathbf{w}^s,\mathbf{w}^f,\boldsymbol{\lambda}) = \int_{\Omega_t^s} \boldsymbol{\lambda}\cdot\left(\mathbf{w}^s - \mathbf{w}^f\right)\: d\Omega.
\end{equation}
The first part of the integral involving the solid test function combined with setting $\mathbf{w}^f=\delta\boldsymbol{\lambda} = \mathbf{0}$ in Equation \ref{eq:fsi_weak_form} reduces to a nodal definition of the Lagrange multiplier,
\begin{equation}
    \lambda_{Ai} = -R_s(N^u_A\mathbf{e}_i,\dot{\mathbf{u}}^s),
    \label{eq:nodal_lagrange}
\end{equation}
where subscript $Ai$ denotes node $A$ vector component $i$ and $\mathbf{e}_{i}$ is the $i$th Cartesian basis vector. This nodal definition can be utilized to eliminate the Lagrange multiplier from the resulting system of equations, reducing the overall size of the linear system and thus the computational cost of the method (see Section \ref{sec:soln_alg}). The second part of the integral reduces to, after substituting in the finite element approximations for each unknown,
\begin{align}
    \int_{\Omega^s_t} \mathbf{w}^f \cdot \boldsymbol{\lambda}\:d\Omega & = \sum_{A\in\omega_v}\sum_{B\in\omega_{\lambda}} (\mathbf{w}^f_A)^T N^v_A(\boldsymbol{\phi}(\mathbf{X}_B,t)) \boldsymbol{\lambda}_B \nonumber \\
    & = \sum_{A\in\omega_v}\sum_{B\in\omega_{\lambda}} (\mathbf{w}^f_A)^T \mathbf{C}^{\lambda}_{AB} \boldsymbol{\lambda}_B
\end{align}
where $\mathbf{C}^{\lambda}_{AB}\in\mathbb{R}^{dim\times dim}$ is given by
\begin{equation}
    \mathbf{C}^{\lambda}_{AB} = N^v_A(\boldsymbol{\phi}(\mathbf{X}_B,t)) \mathbf{I}
    \label{eq:fsi_constraint_matrix}
\end{equation}
where $\mathbf{I}\in\mathbb{R}^{dim\times dim}$ is the identity matrix. The operator $\mathbf{C}^{\lambda}\in\mathbb{R}^{n_{f,dof}\times n_{s,dof}}$ ($n_{i,dof}$ denotes the number of degrees of freedom) represents a mapping from the finite element space defined on the immersed solid to the finite element space defined on the fluid domain, and is commonly referred to as ``force spreading" in the immersed boundary (IB) literature as the solid force (in this case Lagrange multiplier) is ``spread" to the fluid nodes via this operator. In the present work, we refer to this operator as the \textit{FSI constraint matrix}. Here, this operation is performed with the finite element basis functions rather than an approximation of the Dirac delta distribution traditionally used in IB methods to transfer Lagrangian to Eulerian variables. It has been shown in \cite{wang_interpolation_2010} that the use of finite element shape functions for this transfer operation provides a sharper representation of the fluid-solid interface due to the compact support of finite element basis functions, whereas the support of an approximate Dirac delta function typically spans multiple elements.

\begin{remark}
    The choice of finite element basis functions as delta functions at the solid nodes for the Lagrange multiplier finite element space is not the only option, see \cite{hesch_mortar_2014} for a discussion on alternative choices for this space. Our choice in the present work is motivated by computational efficiency.
\end{remark}

\subsection{Solution Algorithm} \label{sec:soln_alg}
The above equations are discretized in time utilizing the implicit second-order accurate generalized-$\alpha$ method \cite{chung_time_1993,jansen_generalized-_2000}. Specifically, we use a variant of the method from \cite{liu_note_2021} that recovers second order accuracy in time for pressure. The resulting nonlinear system of equations to advance the solution from the $n$th time step $t_n$ to the $(n+1)$st time step $t_{n+1}$ are solved using Newton's method and a modified predictor-corrector algorithm presented in \cite{black_immersed_2025}. Here, we briefly summarize the algorithm and state the matrix system solved at each time step, and refer the reader to \cite{black_immersed_2025} for more details. In our approach, we solve a monolithic system at each time step to obtain updates for all unknowns at once. However, due to the special choice of Lagrange multiplier finite element space, we can eliminate the Lagrange multiplier unknowns using the nodal definition of this variable provided by Equation \ref{eq:nodal_lagrange}. Furthermore, due to strong enforcement of the FSI constraint, there exists only one velocity field in our formulation. Consequently, the solid velocity unknowns are related to the fluid velocity unknowns via Equation \ref{eq:nodal_solid_velo} further simplifying the linear system. As a result, we solve the following reduced matrix system for corrections to our predictors within each time step
\begin{equation}
    \begin{bmatrix}
        \frac{\partial\mathbf{R}^{f,m}}{\partial \mathbf{\Dot{V}}_{n+1}} + \mathbf{C}^{\lambda}\frac{\partial\mathbf{R}^s}{\partial \mathbf{\Ddot{U}}_{n+1}}(\mathbf{C}^{\lambda})^T & \quad\frac{\partial\mathbf{R}^{f,m}}{\partial \mathbf{\Dot{P}}_{n+1}}
        \\ \\        
        \frac{\partial\mathbf{R}^{f,c}}{\partial \mathbf{\Dot{V}}_{n+1}} & \quad\frac{\partial\mathbf{R}^{f,c}}{\partial \mathbf{\Dot{P}}_{n+1}}
    \end{bmatrix}
    \begin{bmatrix}
        \Delta \mathbf{\Dot{V}}_{n+1} \\ \\
        \Delta \mathbf{\Dot{P}}_{n+1}
    \end{bmatrix}
    = -
    \begin{bmatrix}
        \mathbf{R}^{f,m} + \mathbf{C}^{\lambda} \mathbf{R}^s \\ \\
        \mathbf{R}^{f,c}
    \end{bmatrix},
\end{equation}
with the increment in the solid acceleration given by
\begin{equation}
    \Delta\Ddot{\mathbf{U}}_{n+1} = (\mathbf{C}^{\lambda})^T\Delta\Dot{\mathbf{V}}_{n+1},
\end{equation}
where $\bold{R}^{f,m}$ is the fluid momentum residual vector obtained by setting $q=0$ in Equation \ref{eq:fluid_weak_form}, $\bold{R}^{f,c}$ is the fluid mass residual vector obtained by setting $\mathbf{w}^f=\mathbf{0}$ in Equation \ref{eq:fluid_weak_form}, $\bold{R}^s$ is the solid momentum residual vector, and $\Dot{()}$ denotes a time derivative. $\bold{V},\bold{P},\bold{U}$ denote degree of freedom vectors for the velocity, pressure, and solid displacement fields respectively.

\begin{remark}
    Note the above matrix system for our immersed FSI method is the same size as that of a fluid-only problem, which of course reduces the overall computational cost of the method. Additionally, due to the 2x2 block structure of the above system, the matrix problem can be solved iteratively using block preconditioning techniques originally developed for incompressible flows \cite{ryan_t_black_computational_2024,deparis_parallel_2014,liu_nested_2020,benzi_numerical_2005,quarteroni_factorization_2000,elman_taxonomy_2008}. Here, we utilize block preconditioners that are constructed directly from the sub-matrices of the block system to improve robustness with respect to changes in application specific parameters.
\end{remark}

\section{Design of the MFEMiFSI Plugin}\label{sec:plugin_design}

\begin{figure}[h]
    \centerline{\includegraphics[width=\columnwidth]{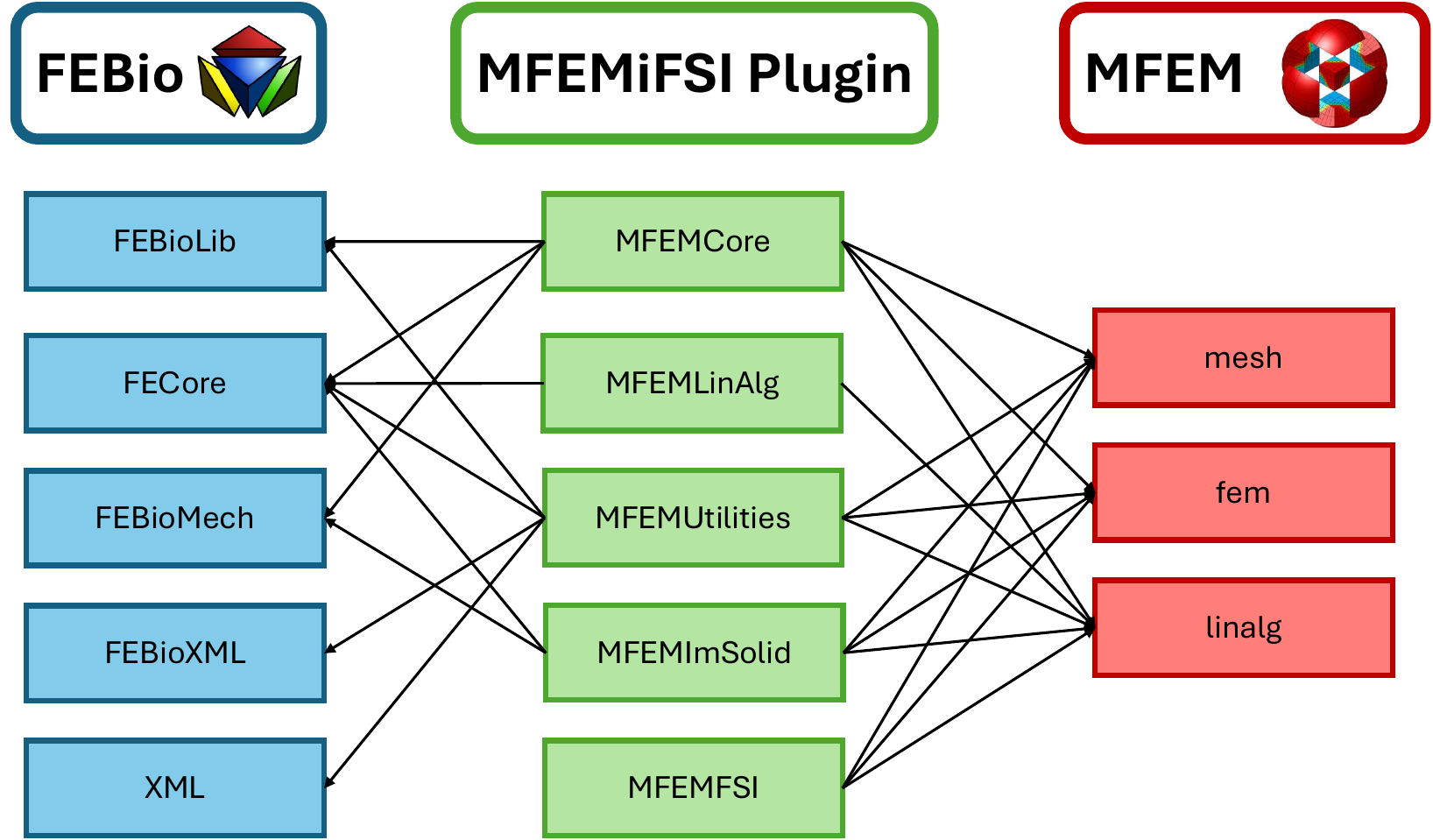}}
    \caption{Diagram of the plugin's design interfacing with FEBio and MFEM.}
    \label{fig:code_diagram}    
\end{figure}

The method described in the previous section has been implemented as an FEBio solver plugin (see \cite{maas_plugin_2018}), termed the MFEMiFSI plugin, which introduces immersed FSI simulation capabilities to FEBio and extends its existing FSI support to problems involving large solid deformations and motions. Additionally, the plugin integrates with the parallel C++ FEM library MFEM for access to an additional suite of FEM algorithms/solvers and in particular through the MFEM library we provide support for distributed memory parallelization of the plugin via the Message Passing Interface (MPI). Although we utilize the solver plugin framework in FEBio, the present work leverages this plugin interface in a unique way to implement the immersed FSI method via coupling a parallel incompressible flow solver written exclusively in the MFEM library with the solid modeling capabilities of FEBio. Since the present FSI approach utilizes a fully-implicit monolithic coupling, this required a new set of classes to facilitate conversion between MFEM and FEBio data structures and allow for construction of the monolithic matrix system using information obtained from both codes. Figure \ref{fig:code_diagram} shows a diagram of the design of the plugin to leverage both MFEM and FEBio as backends to implement the immersed FSI method. We refer the reader to the FEBio \cite{maas_febio_2012} and MFEM \cite{anderson_mfem_2021,andrej_high-performance_2024} libraries for details on their respective directories. The main components of the plugin are as follows:

\begin{itemize}
    \item \textcolor{darkgreen}{\texttt{MFEMCore}}: contains the main driver class of the solver plugin \texttt{MFEMSolver}, as well as classes that setup the finite element model. 
    \item \textcolor{darkgreen}{\texttt{MFEMLinAlg}}: wrappers to convert between the various linear algebra data types available in each library such as vectors and matrices.
    \item \textcolor{darkgreen}{\texttt{MFEMUtilites}}: additional classes to facilitate coupling between the two codes and implement general features for the plugin.
    \item \textcolor{darkgreen}{\texttt{MFEMImSolid}}: contains a driver class for the immersed solid \texttt{MFEMImmersedFEBioSolid}, as well as classes (e.g. \texttt{MFEMImmersedElasticDomain}) to compute the residual vector and tangent matrix.
    \item \textcolor{darkgreen}{\texttt{MFEMFSI}}: the MFEM based parallel flow solver \texttt{IncompressibleFluidSolver} and a lightweight immersed solid implementation in MFEM.
\end{itemize}

\begin{remark}
    Although a partitioned approach would provide a modular implementation, we chose a monolithic approach for the enhanced robustness provided by this type of coupling as our goal is to provide support for modeling of a wide range of biological FSI problems, which are known to cause stability issues (e.g. added-mass instabilities) for standard partitioned coupling methods \cite{forster_artificial_2007}.
\end{remark}

The starting point for our plugin is the \texttt{MFEMSolver} class, which inherits from the \texttt{FESolver} class in FEBio that describes the type of physics problem solved during an analysis step of the FEBio program. This new class interfaces with FEBio predominantly through two overriding functions: \texttt{Init()} and \texttt{SolveStep()}. The former function constructs and partitions the meshes, as well as performs the initial setup of the solvers utilizing data parsed from the input FEBio file. The latter \texttt{SolveStep()} function is called within the main time-stepping loop of FEBio and defines the solution procedure for each time step, namely the predictor-multicorrector algorithm discussed in Section \ref{sec:soln_alg}. Here, we repeatedly assemble and solve the monolithic immersed FSI matrix system. In the current implementation, the global (parallel) matrix system is defined and solved utilizing the linear algebra data structures and routines in MFEM. The fluid and solid contributions to the matrix system are computed from the MFEM flow solver and FEBio immersed solid solver respectively. The key design feature to simplify not only the formation of the FSI matrix system, but also the MPI parallelism of the plugin was to utilize the numbering of the elements and degrees of freedom (DOFs) provided by MFEM's \texttt{ParFiniteElementSpace} class for the FEBio mesh. As a result, we simply call the FEBio assembly routines, with some small changes, to compute the element contributions and subsequently assemble them into the global (parallel) matrix using the standard approach in MFEM.

In addition to the fluid and solid contributions to the matrix system, the present methodology requires (parallel) computation of the FSI constraint matrix $\mathbf{C}^{\lambda}$. Recall from Equation \ref{eq:fsi_constraint_matrix}, each row of the FSI constraint matrix corresponds to a fluid shape function evaluated at each solid node. Consequently, formation of this matrix requires a point search process and a non-standard assembly procedure. In the present work, we extend the existing parallel point search class in MFEM \texttt{FindPtsGSLIB} \cite{mittal_general_2025}, as well as leverage \textit{hypre}'s \texttt{IJ} interface \cite{falgout_conceptual_2006} and MFEM's \texttt{HypreParMatrix} wrapper class to implement the formation of this matrix. \texttt{FindPtsGSLIB} provides an interface to efficiently communicate between MPI ranks where the point search process was started for a collection of local points and MPI ranks where these points were found. \textit{hypre}'s \texttt{IJ} interface allows the user to interact with a distributed sparse matrix using row and column indices in the typical linear-algebra way. Pseudocode for our algorithm to compute $\mathbf{C}^{\lambda}$ is presented in Appendix \ref{ax:secA1}.

Another design feature of the plugin is modularity for different solid formulations and incorporation of additional solid modeling capabilities. In the present implementation, the base class \texttt{MFEMImmersedElasticDomain} that computes the immersed solid contributions provides an easily extendable interface for computing the element residual vectors and tangent matrices via virtual function definitions (see Appendix \ref{ax:secA2}). As an example, we include within the \textcolor{darkgreen}{\texttt{MFEMImSolid}} directory an implementation of a three field element formulation, the mean dilatation method \cite{simo_quasi-incompressible_1991}, via the class \texttt{MFEMImmersed3FieldElasticDomain} that inherits from the aforementioned base class. This three field formulation circumvents the well-known locking phenomena observed when using displacement-only low-order hexahedral and pentahedral elements for modeling nearly and fully incompressible solids.

\section{Numerical Test Problems}\label{sec:tests}
This section details the numerical benchmark and test problems that have been conducted to verify the implementation and demonstrate the capabilities of the immersed FSI framework. Note since FEBio is a three-dimensional (3D) finite element code, all two-dimensional (2D) problems are modeled in 3D with one element in the z direction and all solution components in the z direction set to zero. Unless stated otherwise, all example problems use first-order Lagrange polynomials as basis functions for the finite element approximations. Meshes for the following problems were constructed using FEBio Studio \cite{maas_febio_2012}, Simvascular \cite{updegrove_simvascular_2017}, or GMsh \cite{geuzaine_gmsh_2009}. Numerical simulations reported in this work were performed on both the Respublica supercomputer at the Children's Hospital of Philadelphia (CHOP) and the Anvil supercomputer at Purdue University.  

\subsection{Immersed annular solid in static equilibrium}
In this section, we model the equilibrium configuration of a thick annular cylinder immersed in a rigid prismatic box with a square cross-section containing incompressible fluid, an FSI benchmark problem with an analytical solution from \cite{boffi_hyper-elastic_2008,roy_benchmarking_2015,heltai_fully_2014,griffith_hybrid_2017}. Specifically, we reuse the problem setup and convergence analysis from \cite{black_immersed_2025}.

\begin{figure}[h]
    \centerline{\includegraphics[trim=0 90 0 0, clip, width=1.0\columnwidth]{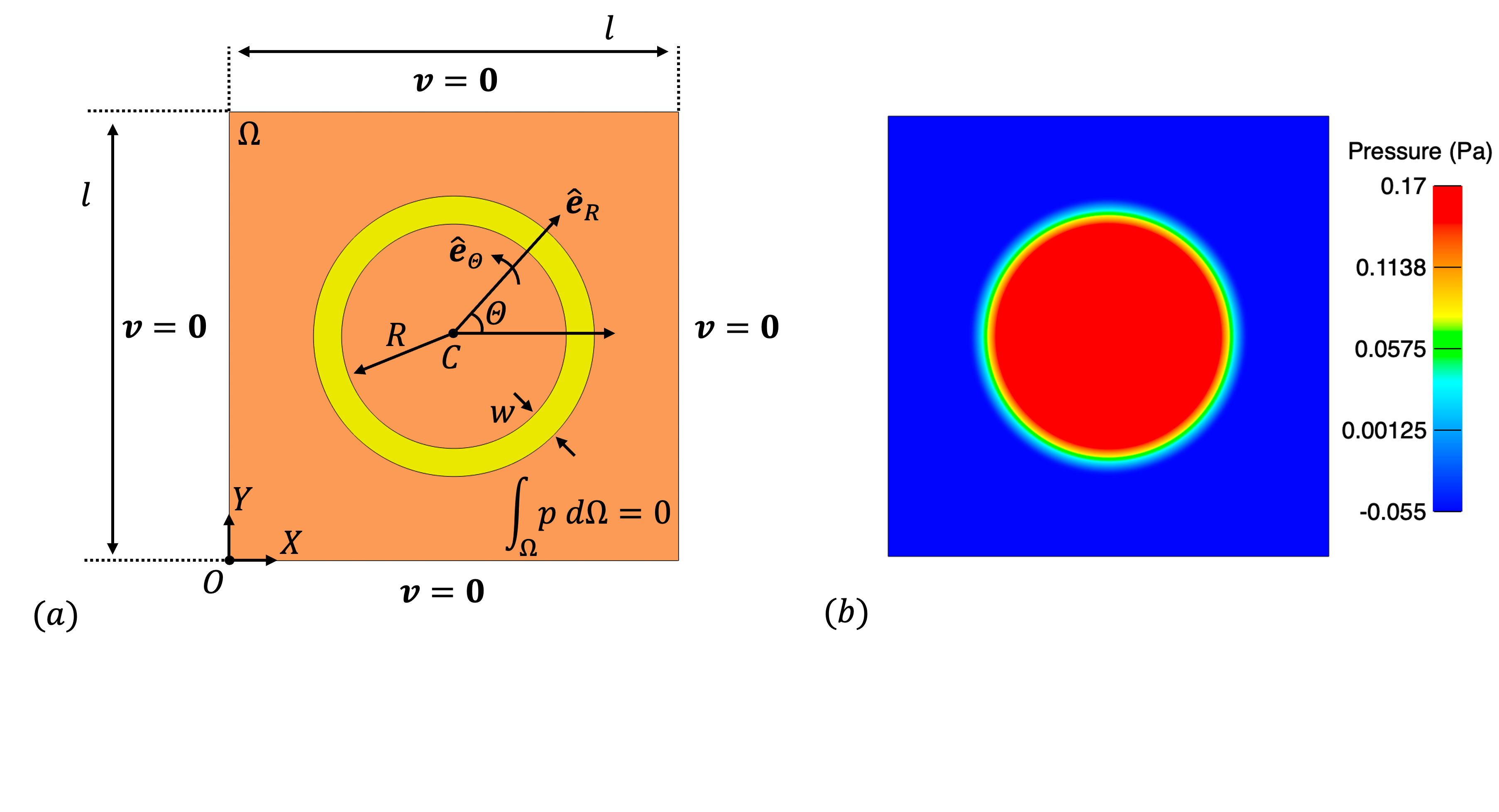}}
    \caption{(a) Problem setup and boundary conditions for the immersed annular solid in static equilibrium test case. The orange region is the fluid domain, while the yellow region is the solid domain. Note that $\hat{e}_{()}$ denotes a unit vector in a coordinate direction. (b) Snapshot of the fluid pressure throughout the domain from the N=256 mesh.}
    \label{fig:im_overview}    
\end{figure}

We consider a two-dimensional setup for this problem, as shown in Figure \ref{fig:im_overview}. A square domain with side length $l = 1\:m$ is filled with incompressible Newtonian fluid and an annular solid with inner radius of $R = 0.25\:m$ and thickness $w = 0.0625\:m$ is immersed within the square domain. The fluid density and dynamic viscosity are $\rho^f = 1.0\:kg/m^3$ and $\mu^f = 1.0\:Pa\cdot s$ respectively. For the solid, we consider a fiber reinforced material with a continuous distribution of concentric fibers that introduce stiffness in the circumferential direction. The 2nd Piola-Kirchoff stress is given by 
\begin{equation}
    \textbf{S}^s = \mu^s \hat{\mathbf{e}}_{\Theta}\otimes\hat{\mathbf{e}}_{\Theta},
\end{equation}
where we consider $\mu^s = 1.0\:Pa$. The initial solid density is set to $\rho_0^s=1.0\:kg/m^3$. Although a relatively simple problem setup and boundary conditions (Figure \ref{fig:im_overview}), it is the internal stresses within the immersed annular solid that generate a non-trivial pressure throughout the domain. We refer the reader to \cite{boffi_hyper-elastic_2008,heltai_fully_2014} for the specific form and a derivation of the analytical solution for this benchmark problem.

\begin{figure}[h]
    \centerline{\includegraphics[width=0.7\columnwidth]{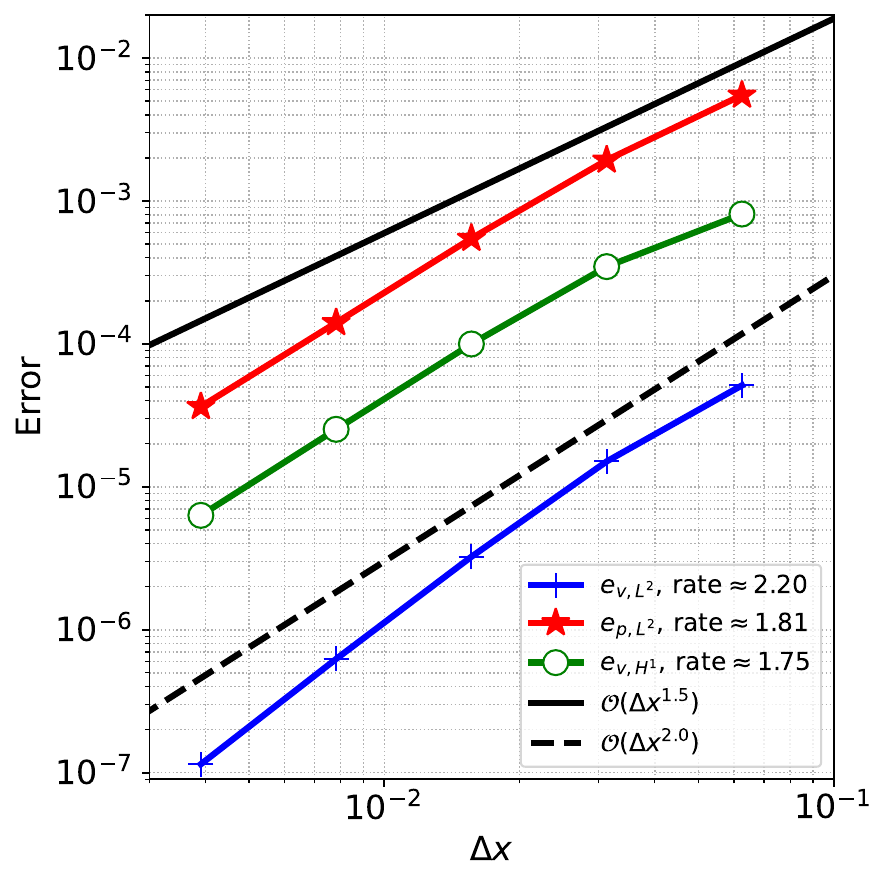}}
    \caption{Convergence results for the immersed annular solid disk in static equilibrium problem.}
    \label{fig:disk_converge}
\end{figure}

For the convergence analysis, we discretize the fluid domain with a uniform Cartesian grid of size $N \times N$, where $N = 16,32,64,128,256$. We consider the solid domain as a rectangle whose short ends are joined together and discretize with $112M \times M$ quadrilateral elements, where $M = 2N/16$ to generate solid meshes with element sizes twice as fine as the fluid mesh size. We simulate 10 time steps of size $\Delta t = 1.0\times10^{-3}$ s to ensure a converged equilibrium configuration is obtained, following \cite{black_immersed_2025}. Figure \ref{fig:im_overview} shows the converged equilibrium fluid pressure from the N=256 mesh. Results of the convergence analysis are shown in Figure \ref{fig:disk_converge}, demonstrating optimal convergence rates of approximately order 2.0 and 1.5 for velocity and pressure in the $L^2$ norm and 1.5 for velocity in the $H^1$ norm, as obtained in \cite{black_immersed_2025}.

\subsection{Idealized heart valve}
The test problems in this section aim to demonstrate the capabilities of the proposed framework to simulate valve-like behavior with two model problems.
\subsubsection{Open valve}\label{subsub:openvalve}
For the first example, we consider a 2D valve model problem that has been previously studied in \cite{black_immersed_2025,kamensky_immersogeometric_2015,hesch_continuum_2012,gil_immersed_2010}. We refer the reader to \cite{black_immersed_2025} for a complete description of the problem geometry, boundary conditions, material properties, etc. that are reproduced in the present work. Here, we briefly summarize the problem. 

We consider a rectangular channel of size $8.0 \times 1.61$ cm with two cantilever beams immersed in the fluid domain $2.0$ cm from the left end that each have a thickness of $0.0212$ cm thick and a height of $0.7$ cm. Note the two beams (or leaflets) are purposefully separated to avoid contact phenomena. No-slip boundary conditions are applied on the top and bottom walls of the channel while the outlet of the domain is traction-free. At the left end, a time-dependent parabolic velocity profile is prescribed to simulate pulsatile flow at a Reynolds number of 110. The fluid is modeled as an incompressible Newtonian fluid with density $\rho^f = 100$ g$/\text{cm}^3$ and dynamic viscosity $\mu^f = 10$ Poise. The leaflets are modeled as a compressible Neo-Hookean material (see \cite{bonet_nonlinear_2016}) with a density of $\rho^s_0 = 100$ g$/\text{cm}^3$, Young's modulus $E = 5.6 \times 10^7$ dyne$/\text{cm}^2$, and Poisson's ratio of $\nu = 0.4$. 

\begin{figure}[h]
    \centerline{\includegraphics[scale=0.8]{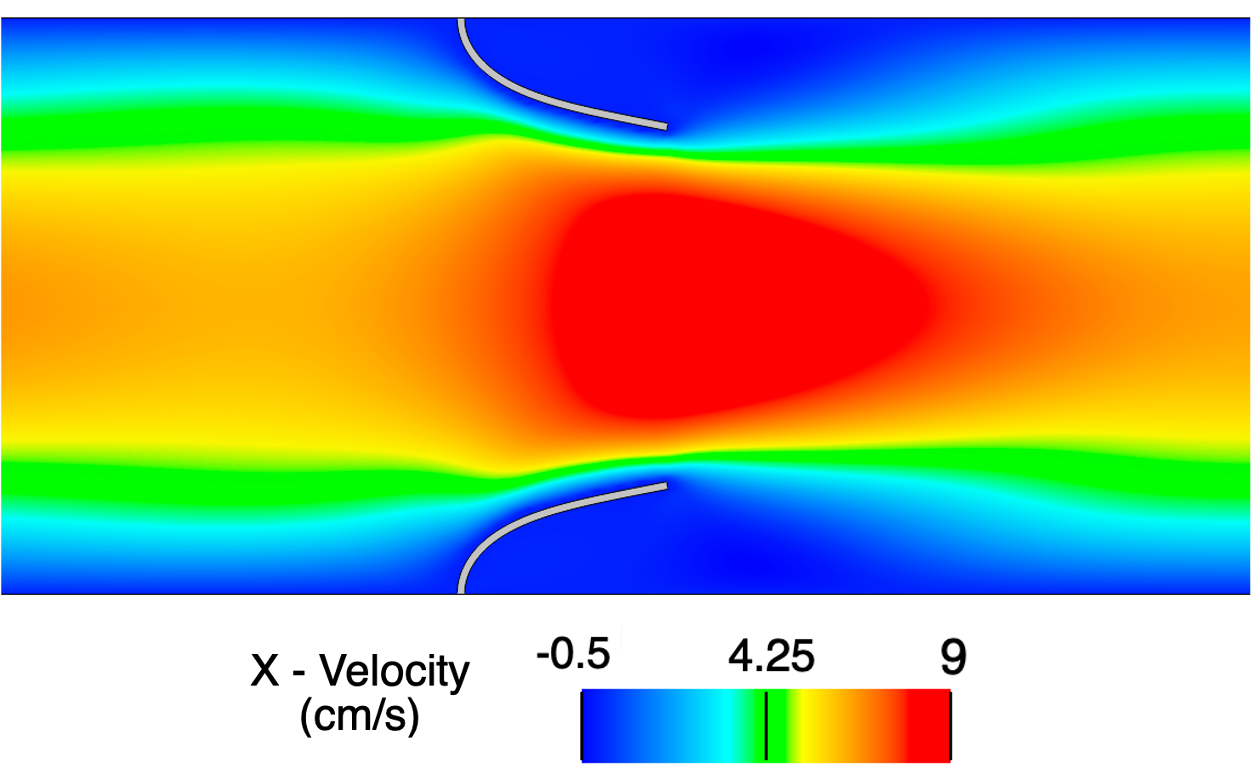}}
    \caption{Snapshot of the numerical solution for the open valve test problem on mesh M3 at time $t=2.25$ s (peak inflow) zoomed in around the leaflets.}
    \label{fig:ov_snapshot}    
\end{figure}

\begin{figure}[h]
    \centerline{\includegraphics[width=1.0\columnwidth]{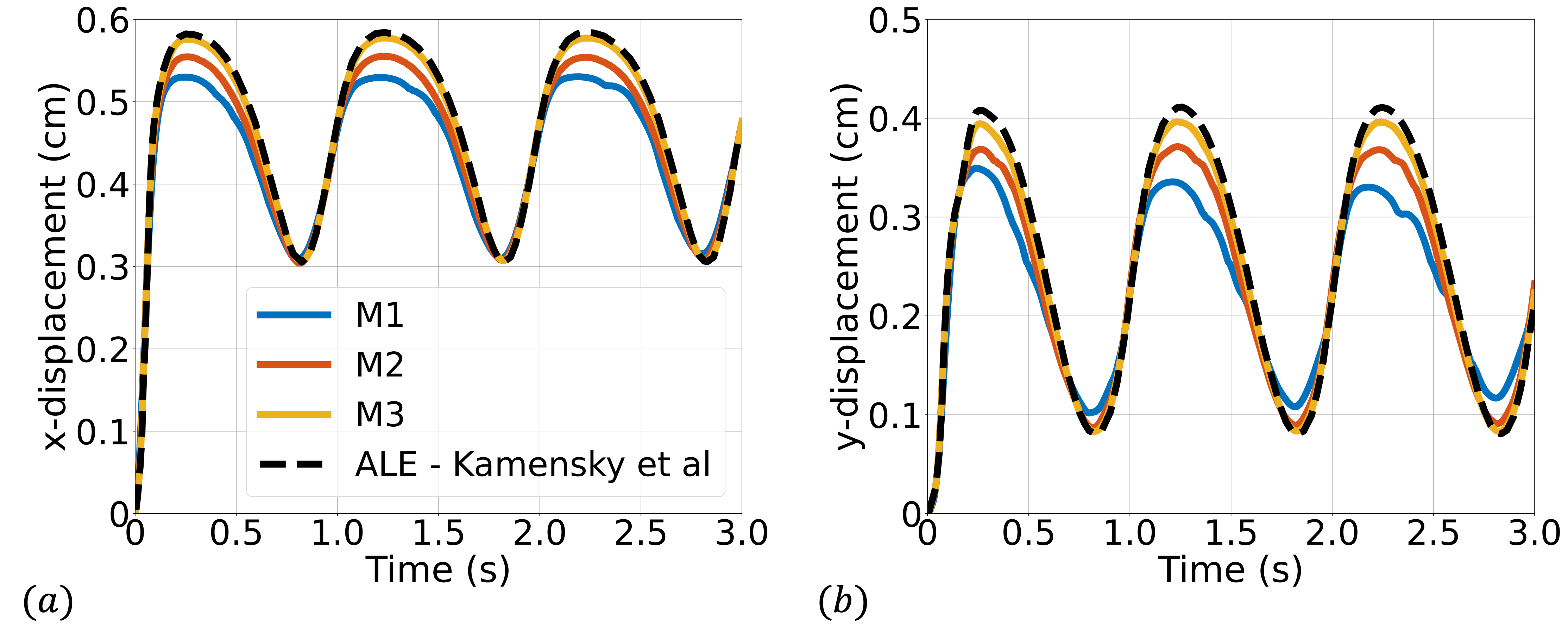}}
    \caption{Time history of the (a) x- and (b) y-displacement of the tip of the top leaflet compared with a reference ALE-FSI simulation from \cite{kamensky_immersogeometric_2015}. M1, M2, and M3 correspond to the sequence of quadrilateral fluid meshes considered of size $129 \times 32$, $256\times 64$, and $512 \times 128$ elements respectively.  \label{fig:hv_open_disp}}  
\end{figure}

Simulations are run to a final time of $3$ s using a time step size of $\Delta t = 1\times10^{-4}$ s. We consider three quadrilateral meshes for the fluid domain (M1 = $129 \times 32$, M2 = $256\times 64$, M3 = $512 \times 128$), while we consider a fixed solid mesh of $4\times100$ quadrilateral elements for each fluid mesh. Figure \ref{fig:ov_snapshot} shows a representative snapshot of the numerical solution at peak inflow. The time history of the x- and y-displacement of the tip of the top leaflet is shown in Figure \ref{fig:hv_open_disp} demonstrating convergence under mesh refinement to a reference boundary-fitted ALE-FSI method from \cite{kamensky_immersogeometric_2015}.

\subsubsection{Closed valve}
In this second example, we consider a benchmark problem that simulates the closed state of a valve (see \cite{boilevin-kayl_numerical_2019}) to assess the capability of our framework to capture pressure discontinuities. The problem setup consists of a rectangular channel of size $4\times 1$ cm with a single elastic beam immersed in the fluid domain whose centerline is located at $2$ cm, has a thickness of $0.0212$ cm, and height of $1$ cm spanning the entire channel in the vertical direction. At the left end, a time-dependent pressure of the form
\begin{equation}
    p(t) = 
    \begin{cases}
        3\times10^5\times\frac{t}{0.1} & \text{if}\quad 0\leq t<0.1 \\
        3\times10^5 & \text{if} \quad t \geq 0.1
    \end{cases},
\end{equation}
is applied. Note that the maximum value of the prescribed pressure is approximately 225 mmHg, which is larger than the normal range of pressures experienced by heart valves \textit{in vivo} (120/80 mmHg) \cite{american_heart_association_understanding_2024}. All other boundary conditions, material properties, etc. are the same as in the first valve model problem (Section \ref{subsub:openvalve}). 

\begin{figure}[h]
    \centerline{\includegraphics[scale=0.5]{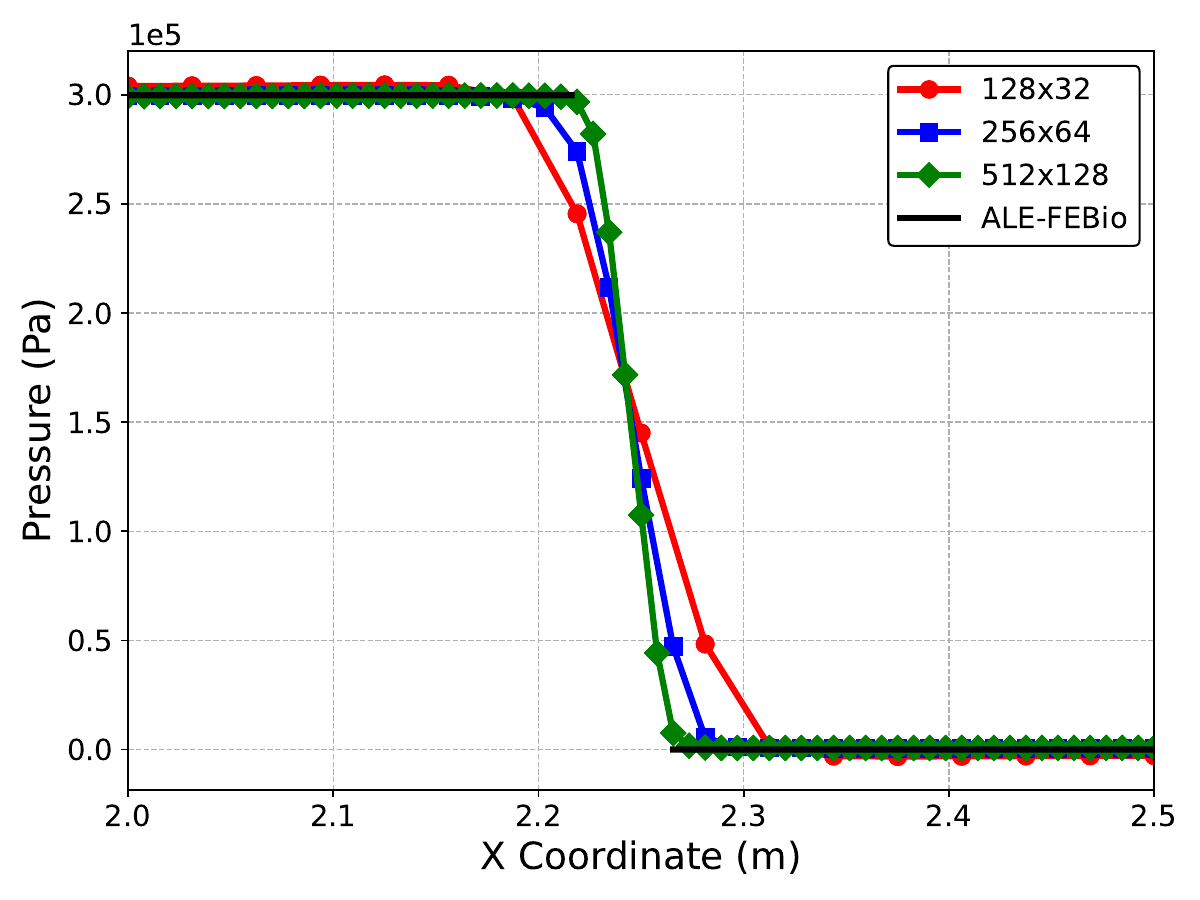}}
    \caption{Centerline fluid pressure for each mesh resolution compared with a reference ALE-FSI calculation using FEBio. Note there exists a gap in the line for FEBio fluid pressure since in the FEBio ALE-FSI formulation fluid pressure is not defined within the solid. Furthermore, the pressure discontinuity in this problem occurs across the finite thickness of the immersed solid.}
    \label{fig:cv_centerline_pressure}    
\end{figure}

\begin{figure*}[t]
\centerline{\includegraphics[width=1.0\textwidth]{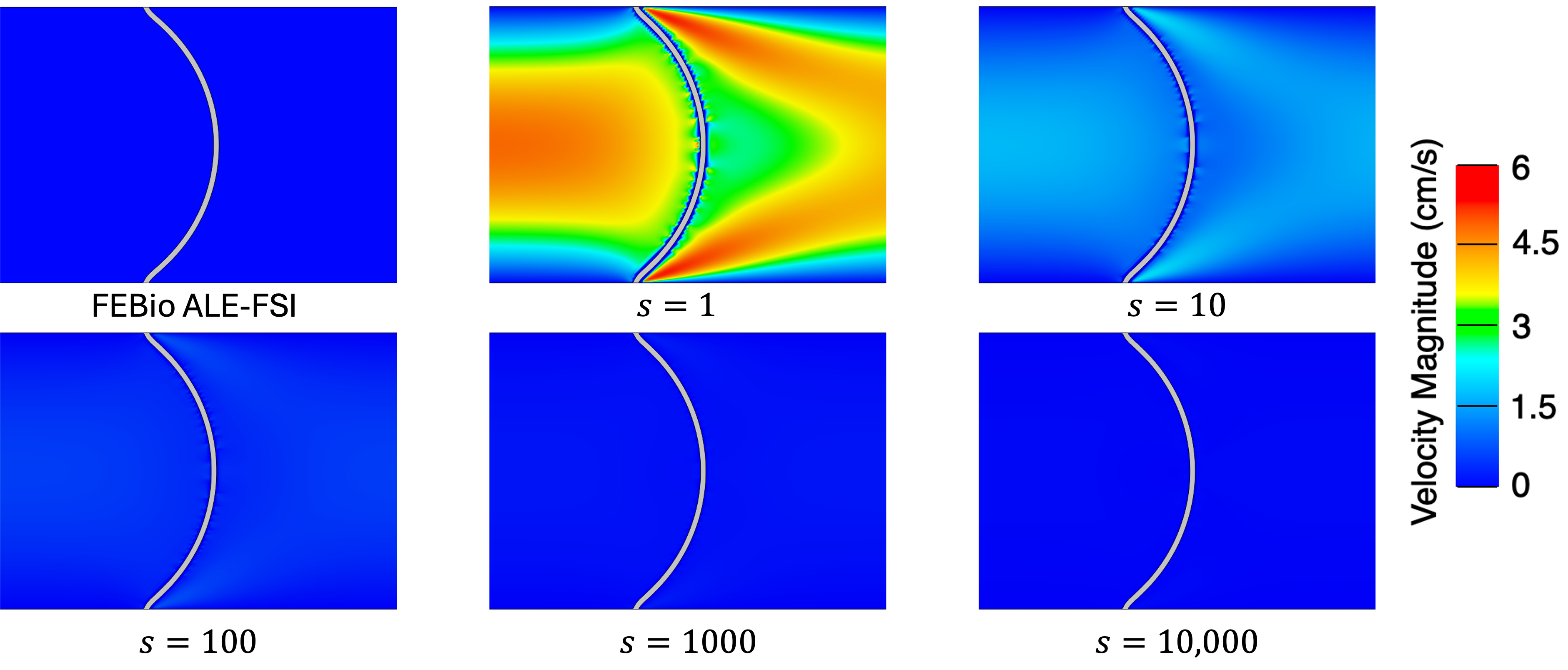}}
\caption{Velocity magnitude near the deformed beam at the final time $t=3$ s for several values of the additional stabilization scaling parameter $s$. \label{fig:cv_flow_varying_s}}
\end{figure*}

We simulate for 3 s to ensure a converged steady-state solution using a time step size of $\Delta t = 1\times10^{-4}$ s. We consider four meshes for the fluid ($128\times32$, $256\times64$, $512\times128$) and corresponding meshes for the solid ($5\times64$, $5\times128$, $5\times256$) to ensure the solid mesh size is twice as fine as the fluid mesh size. Note in this problem, as the pressure ramps up, the solid will deform to the right eventually reaching a steady-state where the pressure on the left and right sides of the beam reach a constant $3\times10^5$ dyne$/\text{cm}^2$ and $0$ dyne$/\text{cm}^2$ respectively. Figure \ref{fig:cv_centerline_pressure} shows the centerline pressure for each mesh size, demonstrating convergence to a reference ALE-FSI solution computed using FEBio on a different mesh with $32,768$ fluid elements and $640$ solid elements \cite{maas_febio_2012,noauthor_febio_nodate,shim_formulation_2019}. Next, we investigate varying the parameter $s$ and to isolate its effect we utilize the finest mesh for subsequent simulations. The steady-state solution to this example problem is a hydrostatic state, i.e. the velocity is zero everywhere. Here, we show that without the modification of the fluid VMS stabilization parameters using the additional scaling factor $s$, the method cannot accurately represent a hydrostatic state (a flow configuration similar to that observed during heart valve closure). As depicted in Figure \ref{fig:cv_flow_varying_s}, there is negligible flow for this example problem when $s>100$.

\subsection{Oscillating flexible leaflet} \label{sec:flexible_leaflet}
In this section, we simulate the motion of a flexible leaflet initially oriented perpendicular to the streamwise direction. Several variants of this problem have been considered previously, see \cite{baaijens_fictitious_2001,yu_dlmfd_2005,kadapa_fictitious_2016,wang_one-field_2017}, and in this work we most closely follow the setup from \cite{wang_one-field_2017}. The purpose of this example problem is to demonstrate the ability of the present formulation to model large and complex solid deformations driven by fluid forces. In particular, this problem would be challenging to model using an ALE-FSI method and would most likely require remeshing or specialized mesh motion algorithms to complete the simulation. 

\begin{figure}[h]
    \centerline{\includegraphics[scale=0.75]{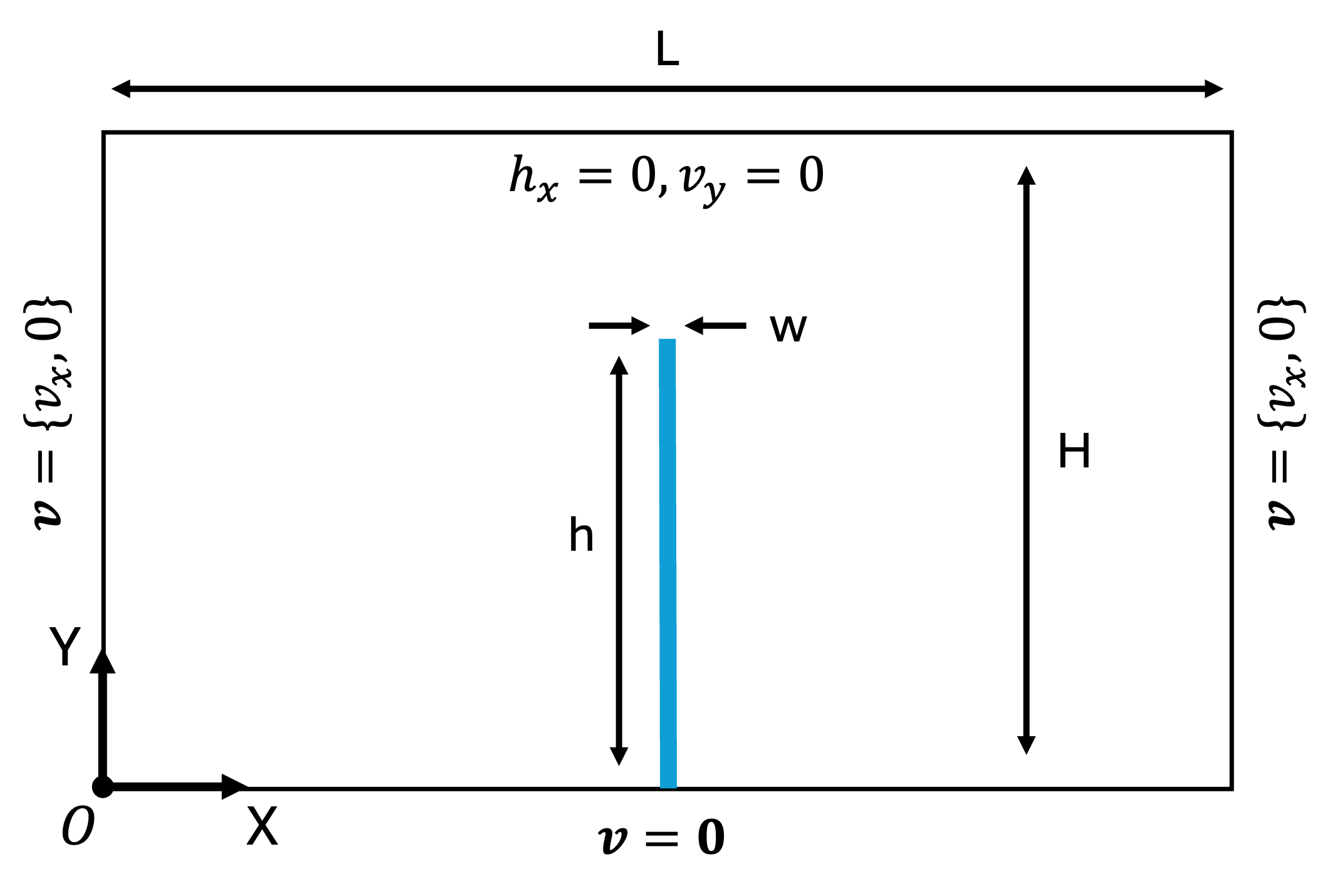}}
    \caption{Problem setup and boundary conditions for the oscillating leaflet example. Note not to scale.}
    \label{fig:os_leaflet_setup}    
\end{figure}

The problem setup and boundary conditions are shown in Figure \ref{fig:os_leaflet_setup}. The fluid channel has length $L = 4.0$ m and height $H = 1.0$ m, while the leaflet has dimensions $w = 0.0212$ m and $h = 0.8$ m. A time-dependent velocity profile is specified at the inlet of the flow domain has the form $\mathbf{v}^f = 15.0y(2-y)\sin(2\pi t) \boldsymbol{\hat{e}}_x$. The fluid is modeled as an incompressible Newtonian fluid with density $\rho^f = 100$ kg$/\text{m}^3$ and dynamic viscosity $\mu^f = 10$ N$\cdot$s$/\text{m}^2$. The solid is modeled as a nearly-incompressible Neo-Hookean material
\begin{equation}
    \psi = \frac{C_0}{2}(\bar{I}_1 - 3) + \frac{\kappa^s}{2}\left( \frac{1}{2}(J^2-1) - \text{ln}J  \right),
    \label{eq:inc_neoHk}
\end{equation} where $C_0 = 1\times10^7$ Pa, $\kappa^s=C_0/10$, and $\bar{I}_1 = \text{tr}\bar{C}$ ($\bar{C}=\bar{F}^T\bar{F}, \bar{F}=J^{1/3}F$). The solid density is set to equal the fluid density $\rho^s = \rho^f$ for this test problem.

\begin{figure}[h]
    \centerline{\includegraphics[width=1.0\columnwidth]{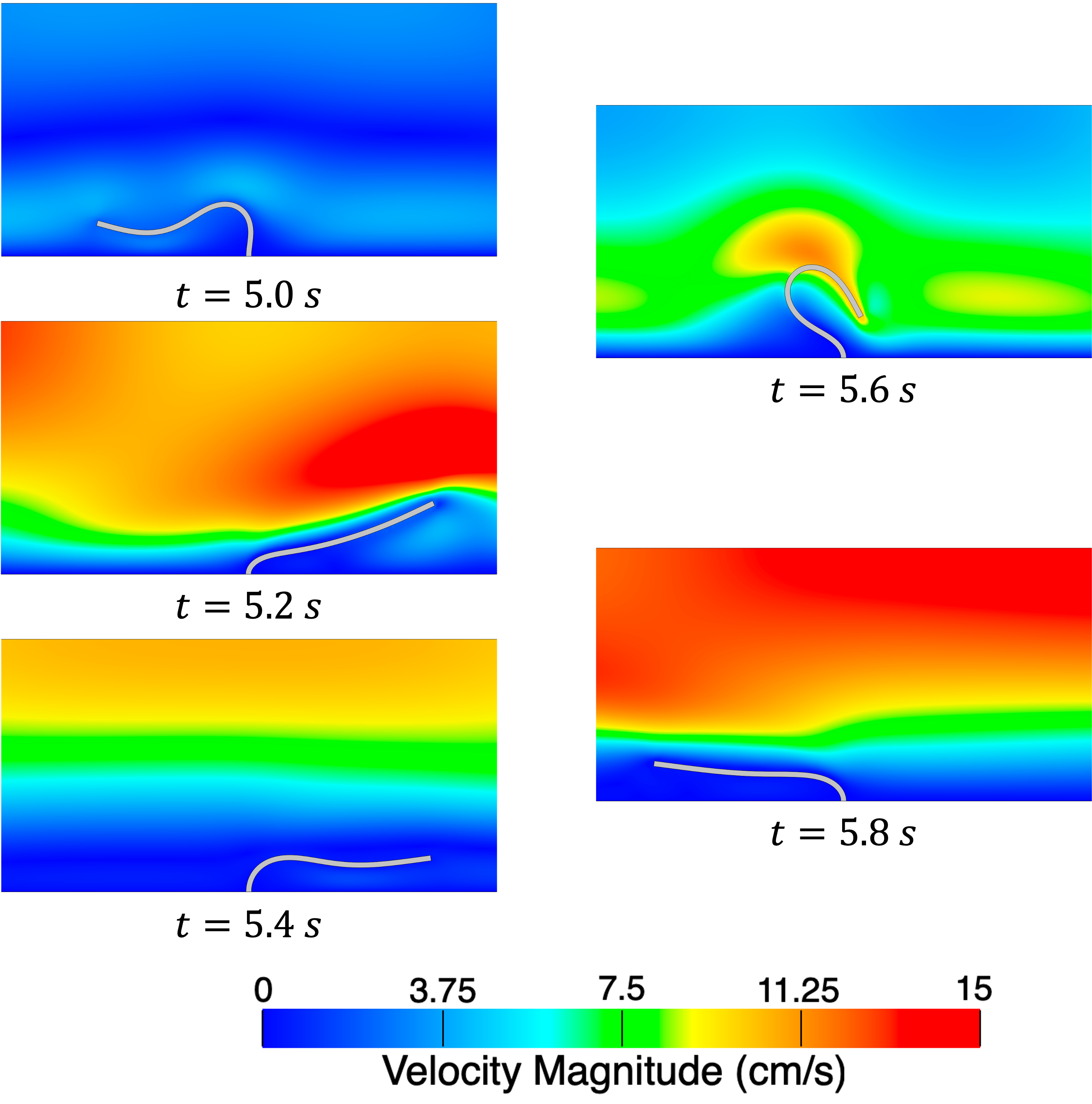}}
    \caption{Snapshots of the velocity magnitude in the fluid domain and leaflet position for the finest mesh at evenly spaced time intervals during the final cycle.}
    \label{fig:os_snapshots}    
\end{figure}

\begin{figure}[h]
    \centerline{\includegraphics[width=1.0\columnwidth]{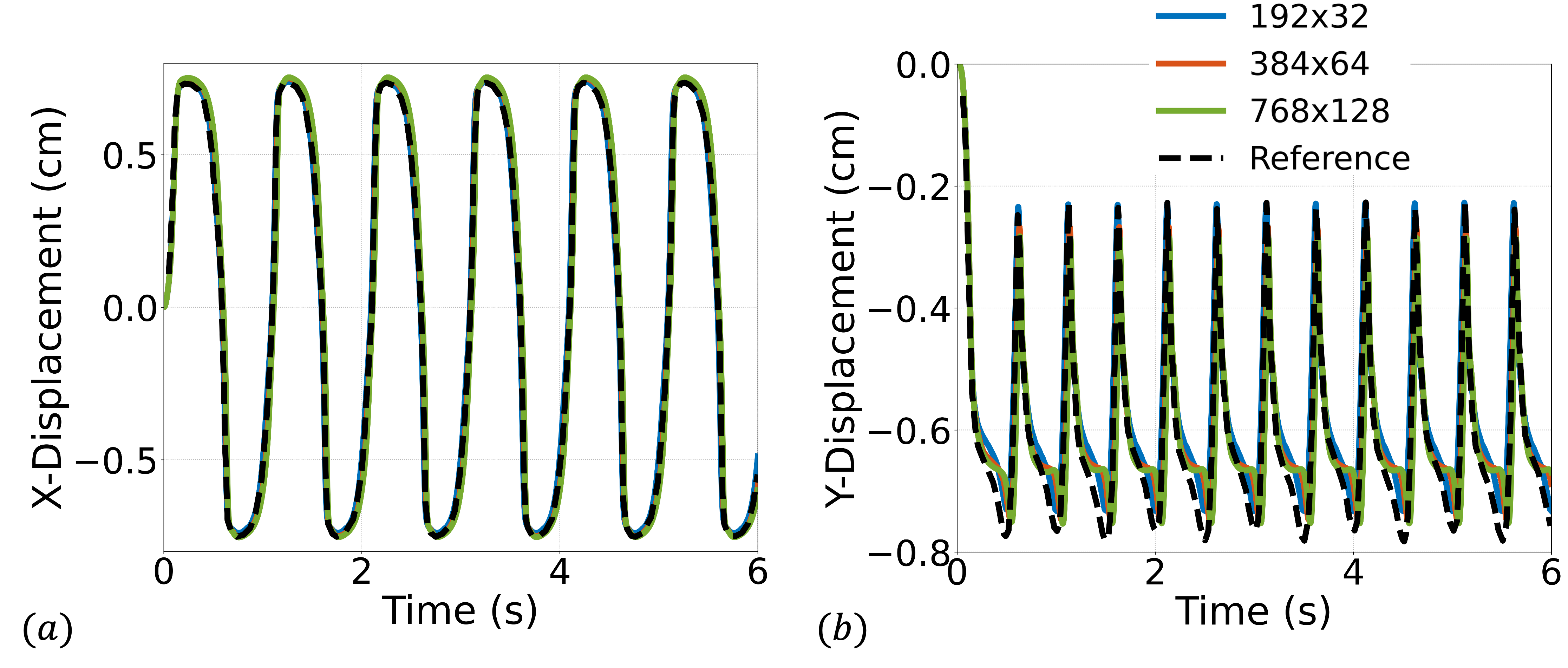}}
    \caption{Time history of the (a) x- and (b) y-displacement of the tip of the leaflet.}
    \label{fig:os_tip_disp}    
\end{figure}

We consider a set of three meshes for the fluid domain $192\times32$, $384\times64$, and $768\times128$ with corresponding solid meshes $4\times45$, $4\times90$, and $6\times180$ respectively. We simulate for 3 s using a time step size of $\Delta t = 1\times10^{-4}$ s, except for the finest mesh where we use $\Delta t = 5\times10^{-5}$ s for numerical stability. Several snapshots of the numerical solution over the final cycle are shown in Figure \ref{fig:os_snapshots}. Time histories of the leaflet tip displacement are shown in Figure \ref{fig:os_tip_disp}. Overall, there is good agreement in the leaflet tip displacement between the present work and the reference solution, with some differences in the peak y-displacements that can be attributed to the use of different elements and different forms of solid constitutive models (uncoupled Neo-Hookean model in the present work). Furthermore, the authors from the reference solution utilized adaptive meshing techniques, making it difficult to reproduce the mesh resolution in the present work. 

\subsection{Free-falling sphere}
In this example, we simulate a free-falling sphere considered in \cite{casquero_nurbs-based_2015}. A free-falling sphere in a fluid reaches its terminal velocity when the downward gravitational force experienced by the object is balanced by the buoyancy and drag force. For a rigid sphere and under the assumption of creeping flow (Re $<<1$), the terminal velocity of a sphere in an unbounded medium is given by (see \cite{casquero_nurbs-based_2015,stokes_effect_1851})
\begin{equation}
    v_t = \frac{2g}{9\mu^f}a^2(\rho^s - \rho^f),
\end{equation}
where $g$ is the gravitational constant and $a$ is the radius of the sphere. Since numerical simulations require a finite domain, we consider a correction factor (see \cite{bohlin_drag_1960}) for the presence of rigid walls of the form
\begin{equation}
    v_{tc} = \frac{v_t}{K}, 
\end{equation}
where we denote the corrected terminal velocity as $v_{tc}$. For a cylindrical domain of radius $A$ and $a/A < 0.6$, the correction factor is given by
\begin{equation}
    K = \left [ 1 - 2.10443\left(\frac{a}{A}\right) + 2.08877\left( \frac{a}{A} \right)^{-3} \right]^{-1}.
\end{equation}
Note that the correction factor decreases the expected terminal velocity in a finite domain since the presence of rigid walls will increase viscous dissipation, which decreases the speed of the free-falling sphere.

\begin{figure}[h]
    \centerline{\includegraphics[width=0.8\columnwidth]{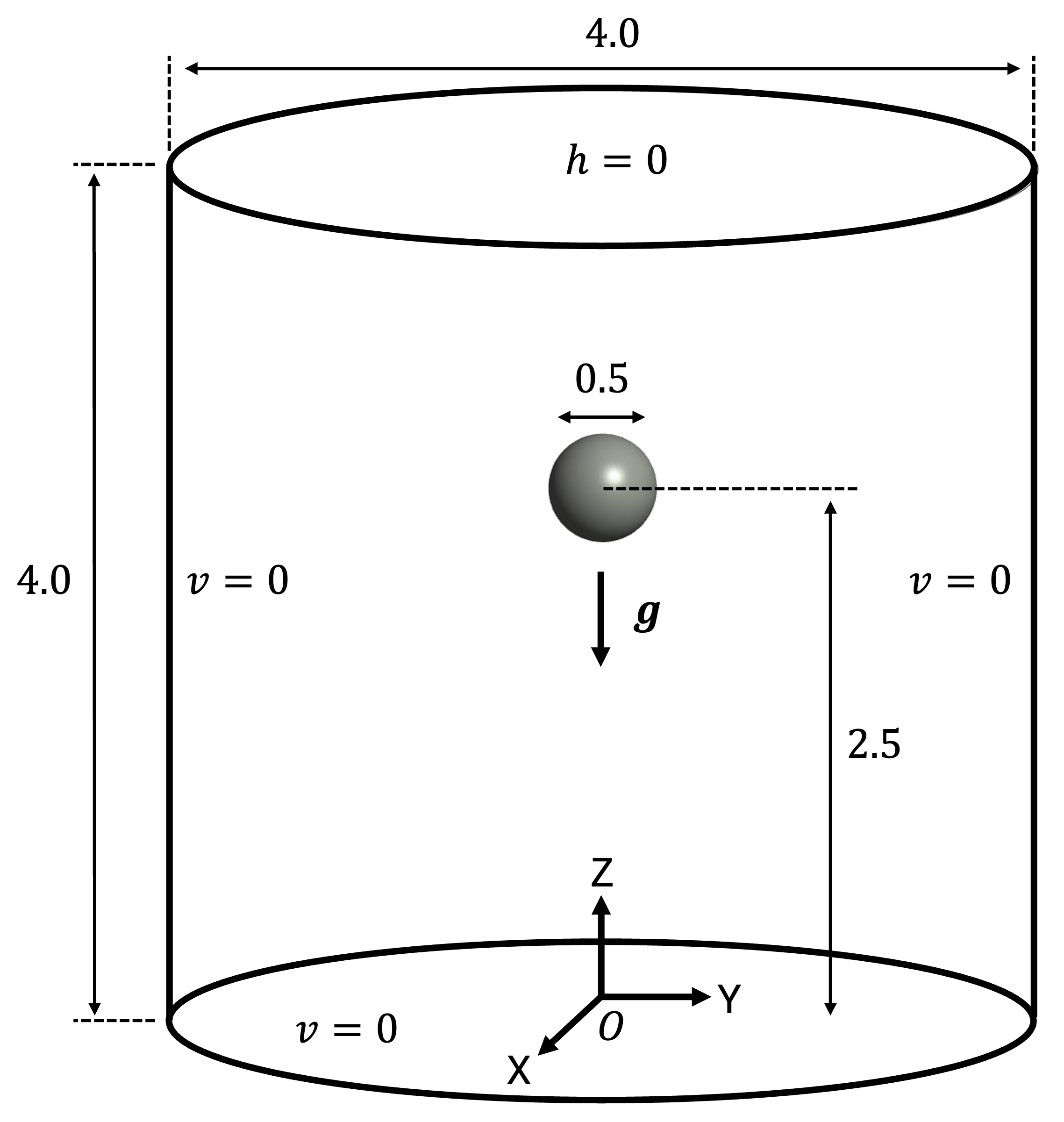}}
    \caption{Problem setup and boundary conditions for the free-falling sphere example.}
    \label{fig:sphere_setup}    
\end{figure}

The problem setup and boundary conditions for the free-falling sphere example are shown in Figure \ref{fig:sphere_setup}. We consider a cylindrical fluid domain of radius $A=2.0$ cm and height 4.0 cm filled with incompressible Newtonian fluid with density $\rho^f = 1$ g$/\text{cm}^3$ and dynamic viscosity $\mu^f = 10$ dyne$/\text{cm}^2$. The cylinder open at the top, where we apply a traction-free condition, while all other boundaries are modeled as no-slip rigid walls. The motion of the sphere is driven by a gravitational force $\mathbf{g} = \{0,0,-981\}$ cm$/\text{s}^2$. A sphere of radius $a=0.25$ cm is placed within the fluid domain such that its center is located at $(0,0,2.5)$ cm to ensure sufficient domain length in the gravitational direction for the sphere to reach its terminal velocity. The solid sphere has an initial density $\rho^s_0 = 1.5$ g$/\text{cm}^3$ and is modeled with a nearly-incompressible Neo-Hookean material (see Section \ref{sec:flexible_leaflet}, Equation \ref{eq:inc_neoHk}) where we set the shear modulus $C_0 = 33550$ dyne$/\text{cm}^2$ sufficiently large to ensure negligible deformation throughout the simulation. 

\begin{figure}[h]
    \centerline{\includegraphics[width=0.8\columnwidth]{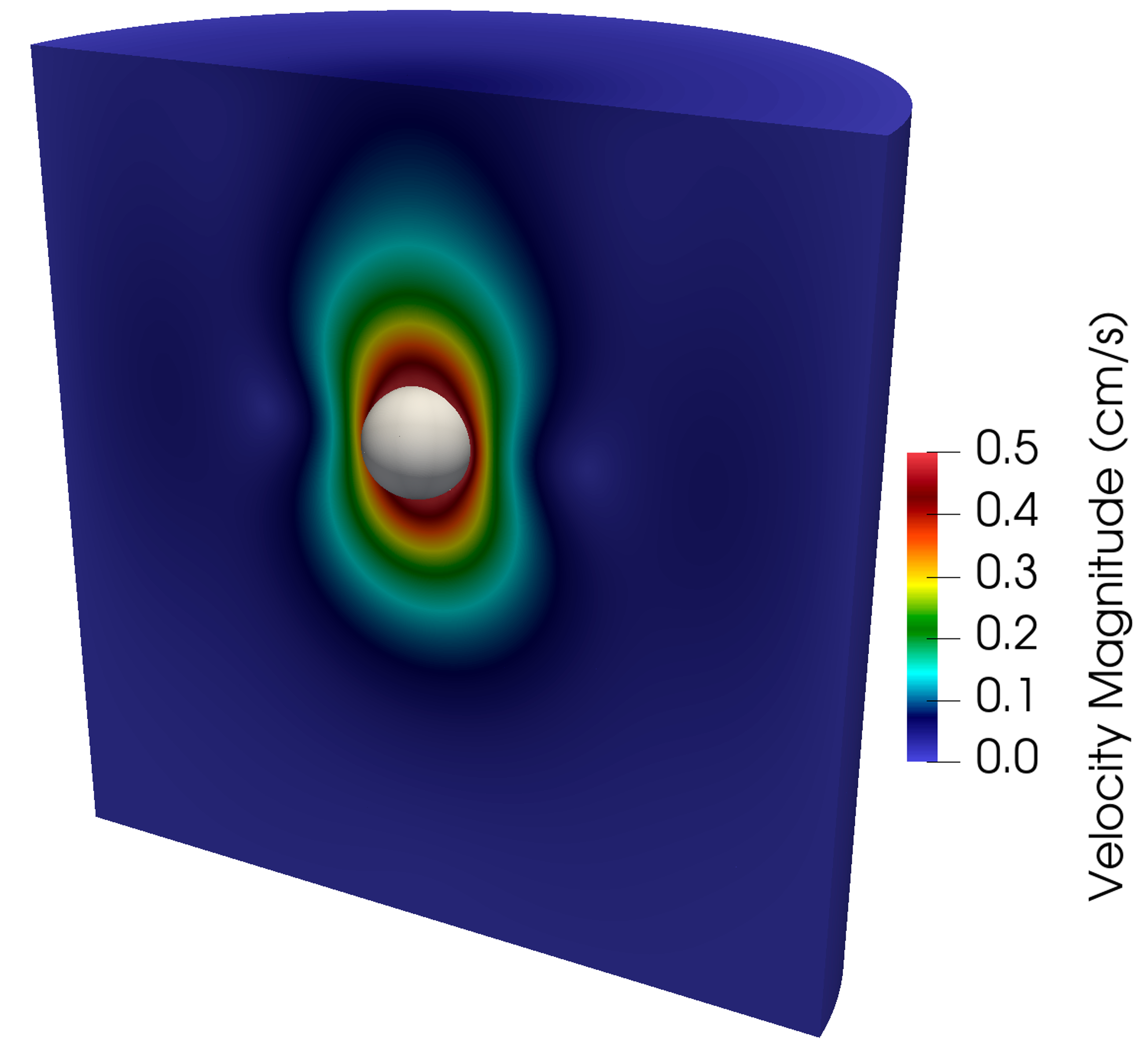}}
    \caption{Snapshot of the velocity magnitude in the fluid domain along with the position of the free-falling sphere shown in gray for the finest mesh considered at $t = 0.1$ s.}
    \label{fig:sphere_velomag}    
\end{figure}

\begin{figure}[h]
    \centerline{\includegraphics[width=0.7\columnwidth]{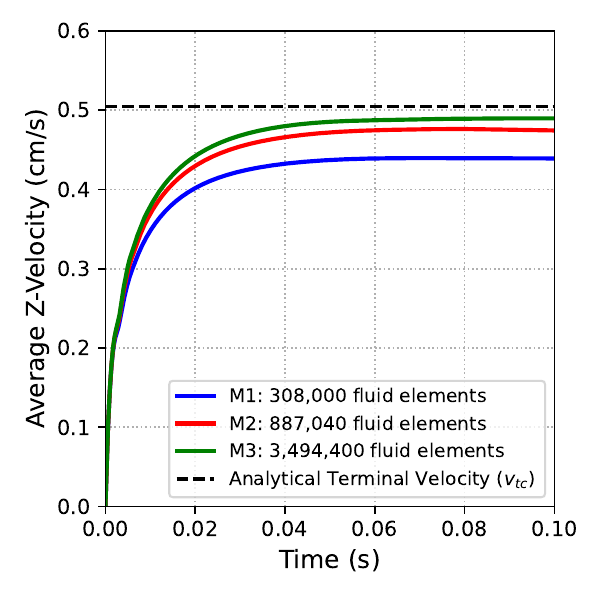}}
    \caption{Time history of the average sphere velocity in the gravitational direction under mesh refinement. Further details on the sequence of meshes utilized in this example problem are provided in Table \ref{tab:freefallsphere_meshes}. Note the dashed black line denotes the analytical solution for a rigid free-falling sphere with the correction factor for the presence of rigid walls in our simulation.}
    \label{fig:sphere_velo}    
\end{figure}

We consider a sequence of three meshes detailed in Table \ref{tab:freefallsphere_meshes}. Simulations are run for $0.1$ s using a time step size of $\Delta t = 1\times10^{-4}$ s, except for the finest mesh where we use $\Delta t=1\times10^{-5}$ s for numerical stability. Figure \ref{fig:sphere_velomag} shows a snapshot of the fluid velocity magnitude and position of the free-falling sphere at $t=0.1$ s on the finest mesh. Time histories of the average sphere velocity in the gravitational direction for each mesh are shown in Figure \ref{fig:sphere_velo} with the reference analytical solution for a rigid sphere in an infinite domain denoted by the black dashed line. Note the convergence under mesh refinement towards the analytical solution, and in particular the percent difference between the numerical solution on the finest mesh and the analytical solution is $3.02\%$. Of course we could increase the stiffness of the sphere to better approximate the rigid (infinite stiffness) sphere considered in the analytical solution, but this can lead to an ill-conditioned matrix.

\begin{table}[htbp]
\caption{Details of the meshes utilized in the free-falling sphere problem. $N_{el}$ corresponds to the number of elements in the mesh, $h$ corresponds to element size and DOFs are the number of degrees of freedom.}
\label{tab:freefallsphere_meshes}
\centering
\begin{tabular}{lllll c c c c c}
\toprule
\textup{Mesh}  & Fluid $N_{el}$ & Fluid DOFs & Solid $N_{el}$  & Solid DOFs \\
\midrule
M1         & 308,000 & 1,274,796 & 23,328 & 73,119 \\
M2         & 887,040 & 3,628,716 & 157,216 & 482,463 \\
M3         & 3,494,400 & 14,173,796 & 1,048,576 & 3,183,363 \\
\bottomrule
\end{tabular}
\end{table}

\subsection{Semilunar heart valve} 
\begin{figure}[h]
    \centerline{\includegraphics[width=1.0\columnwidth]{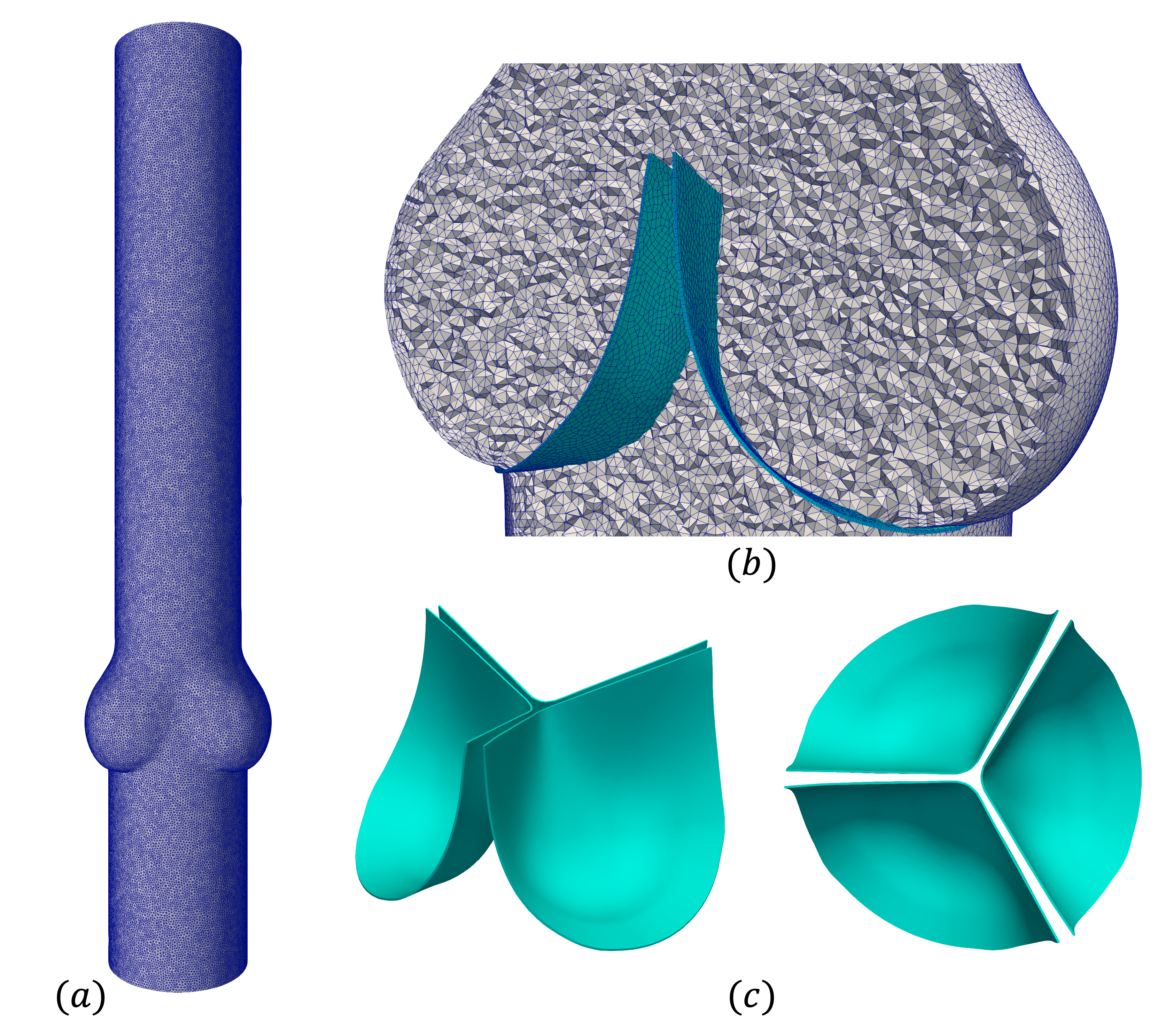}}
    \caption{(a) Computational mesh for the fluid domain, with flow from bottom (inlet) to top (outlet). (b) Cross-section view of the fluid mesh in gray and valve mesh in turquoise within the idealized aortic root. (c) Isometric and top views of the valve geometry.}
    \label{fig:valve_fsi_mesh}    
\end{figure}

In this section, we demonstrate the capability of the present framework to perform FSI simulation of three-dimensional heart valves immersed within a rigid blood vessel. Here, we simulate a tri-leaflet semilunar heart valve within an idealized aortic root. The present geometries, consisting of the leaflets and aortic root shown in Figure \ref{fig:valve_fsi_mesh}, are modifications of the model provided from \cite{ansys_inc_ls-dyna_nodate}. The fluid domain is $18.2$ cm in length and has a diameter of $2.8$ cm, as well as a three-lobed region near the valve to model the aortic root. The inlet, at the bottom of the geometry in Figure \ref{fig:valve_fsi_mesh}, is $4$ cm upstream of the root, while the outlet is $11.8$ cm downstream of the root. The fluid domain was meshed with $8,565,387$ tetrahedral elements, including boundary layer refinement near the vessel wall as seen in Figure \ref{fig:valve_fsi_mesh}, resulting in a total of $5,593,396$ DOFs for both the fluid velocity and pressure approximations. The leaflets have a height of $1.6$ cm and thickness of $0.015$ cm, and were meshed using a combination of hexahedral and pentahedral elements with four elements across the thickness, resulting in $63,452$ elements and $215,970$ solid displacement DOFs.

\begin{figure}[h]
    \centerline{\includegraphics[width=0.7\columnwidth]{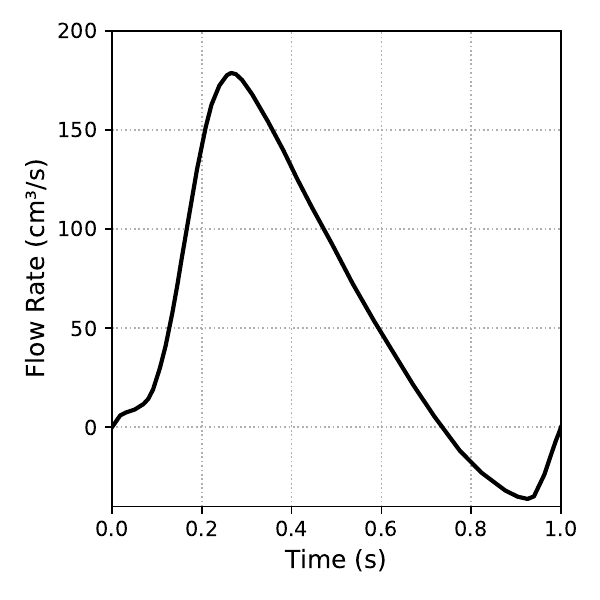}}
    \caption{Flow rate applied to the inlet of the computational domain using a plug flow velocity profile from \cite{oks_fluidstructure_2022}.}
    \label{fig:valve_flow_rate}    
\end{figure}

The fluid is modeled as an incompressible Newtonian fluid with density $\rho^f=1.1$ g$/$$\text{cm}^3$ and dynamic viscosity $\mu^f = 0.036$ dyn$\cdot$s$/\text{cm}^2$. The heart valve is modeled as a compressible Neo-Hookean material (see \cite{bonet_nonlinear_2016}) with density $\rho^s_0 = 1$ g$/\text{cm}^3$, Young's modulus $E = 2\times10^6$ $\text{dyne}/\text{cm}^2$, and Poisson's ratio of $\nu = 0.4$.  At the inlet of the fluid domain, a time-dependent velocity of the form $\mathbf{v}^f = -(Q/A)\mathbf{n}$ is prescribed, where $Q$ is the flow rate shown in Figure \ref{fig:valve_flow_rate}, $A$ is the area of the inlet surface, and $\mathbf{n}$ is the unit normal of the inlet surface. No-slip boundary conditions are applied to the sides of the vessel, while the outlet of the domain is traction-free. To seal the gap between the leaflet and vessel wall, the edges of the leaflets are extended to protrude beyond the vessel geometry. Recall we utilize an immersed FSI approach that does not require the solid and fluid domains to perfectly align. This problem setup is based on the aortic valve FSI problem in \cite{oks_fluidstructure_2022}. Since there is no flow entering the domain at $t=0$, we begin the simulation from an initial stress-free and stationary state. We simulate a single heartbeat given by one cycle of the prescribed flow rate at the inlet of the domain, i.e. total simulation time of $1$ s, using a time step size $\Delta t = 1\times10^{-4}$ s.

\begin{figure}[h]
    \centerline{\includegraphics[width=1.0\columnwidth]{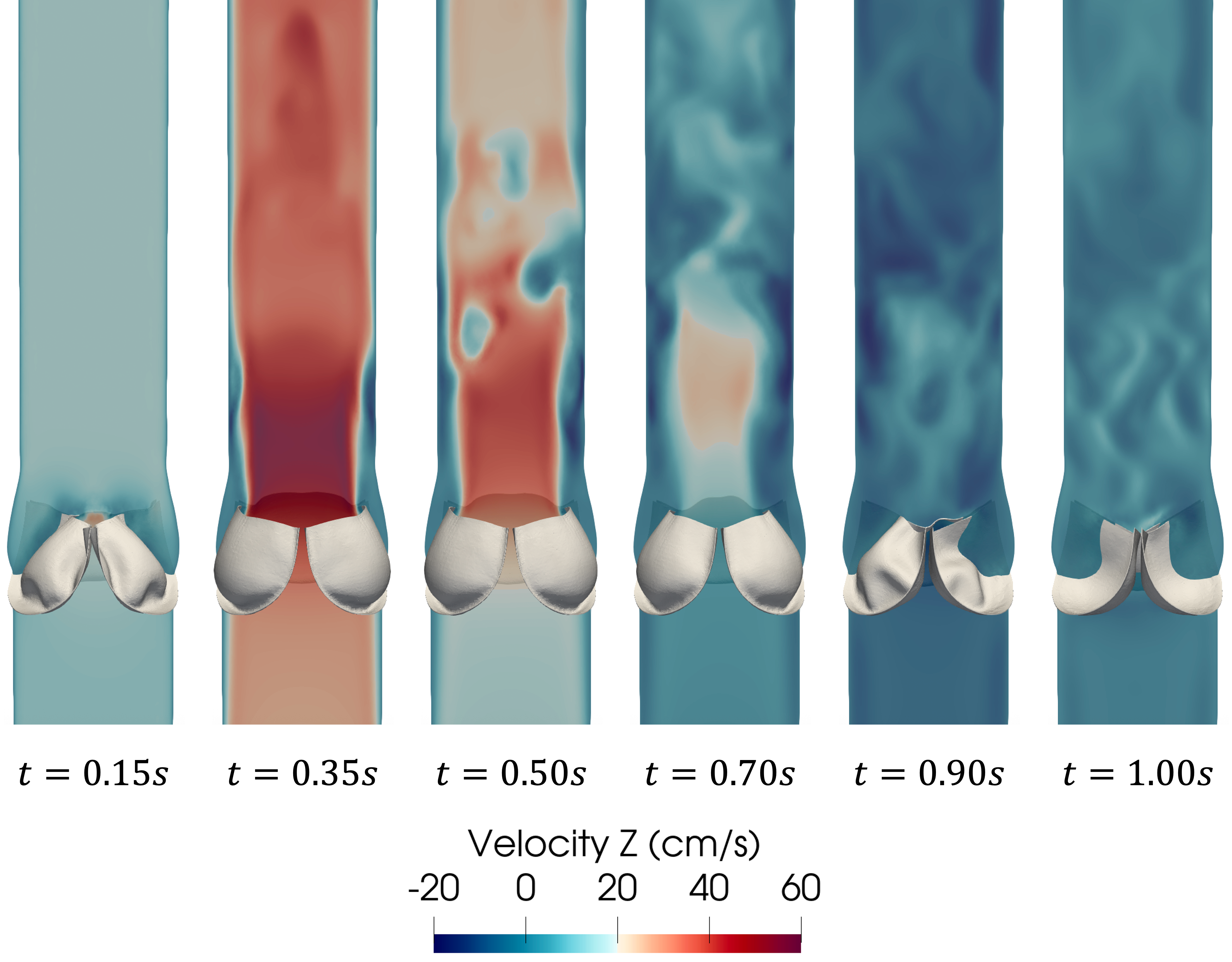}}
    \caption{Snapshots of the z-component of the fluid velocity (component parallel to the centerline of the vessel) plotted on a vertical cross-section and leaflet motion at several time points over the cardiac cycle.}
    \label{fig:valve_fsi_velo}    
\end{figure}

\begin{figure}[h]
    \centerline{\includegraphics[width=1.0\columnwidth]{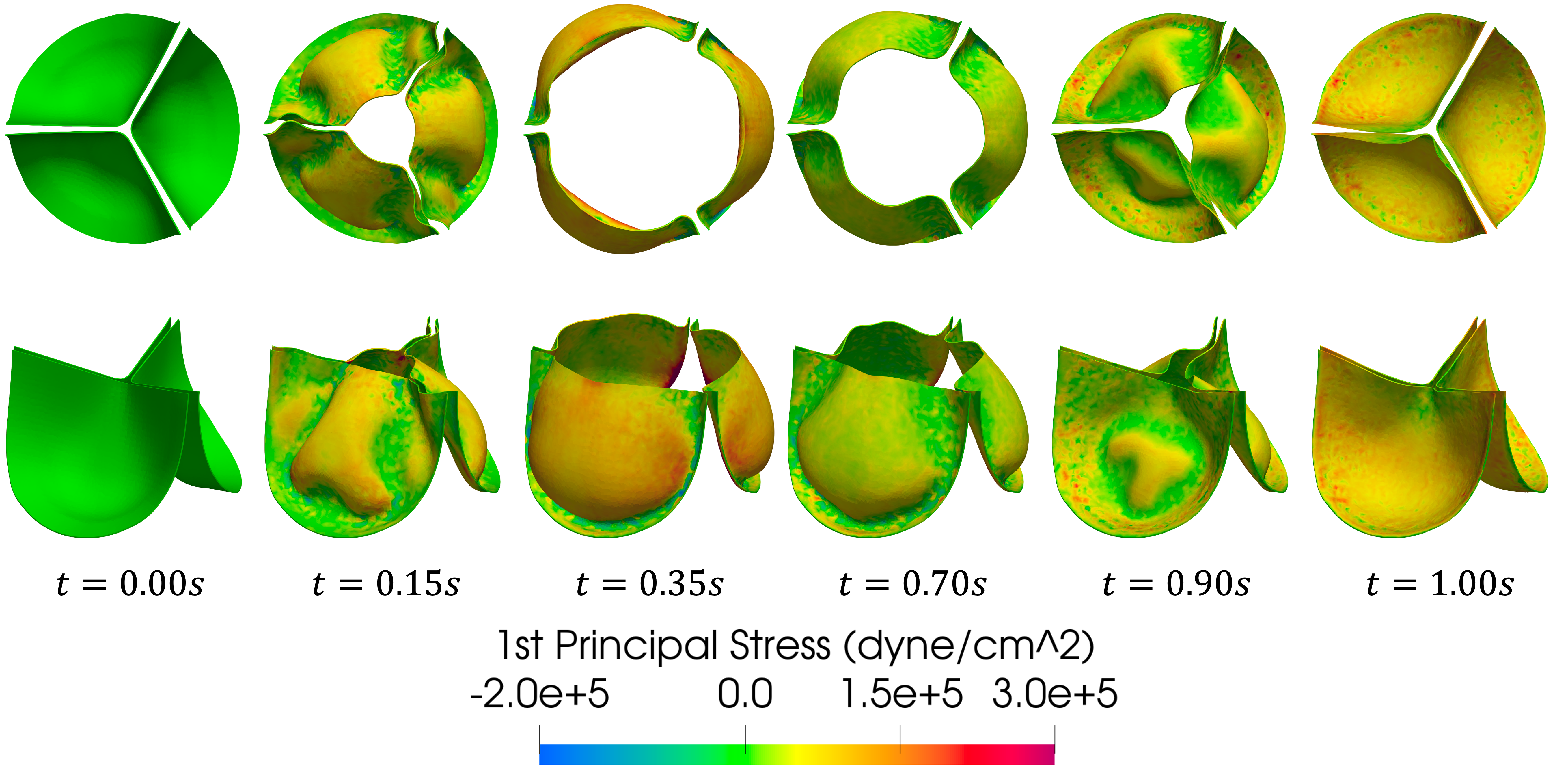}}
    \caption{Snapshots of the leaflet 1st principal stress at several instances over the simulated heartbeat.}
    \label{fig:valve_fsi_disp}    
\end{figure}

Figure \ref{fig:valve_fsi_velo} shows a series of snapshots of the leaflet motion and z-component of the fluid velocity over the simulated heartbeat. The flow field shows the development of a jet upon valve opening, followed by breakdown of the jet as the valve closes until forward flow essentially stops upon valve closure. Note the complex vortex dynamics downstream of the valve, which are expected flow features given the peak Reynolds number is approximately $5,400$ computed using the vessel diameter and peak velocity magnitude. We refer the reader to \cite{bornemann_leaflet_2025,nitti_numerical_2022} for discussions on turbulent flow features downstream of the valve. A top and isometric view of the leaflet 1st principal stress over the cardiac cycle is shown in Figure \ref{fig:valve_fsi_disp}. Note the nonsymmetric deformations during valve closure, which are most likely attributed to the downstream nonsymmetric vortex dynamics. This asymmetric behavior is consistent with valve FSI simulations reported in \cite{oks_fluidstructure_2022,kamensky_immersogeometric_2015,borazjani_fluidstructure_2013}.

\section{Discussion}\label{sec:discussion}
We present a novel open-source immersed FSI framework that addresses critical gaps in publicly available computational biomechanics software through three principal contributions: (1) coupling of the GPU-ready, scalable MFEM library with FEBio's sophisticated biomechanics modeling capabilities, thereby extending FEBio's FSI modeling support to include large solid deformations via an immersed formulation; (2) introduction of an incompressible flow solver to FEBio with support for distributed memory parallelization; and (3) a modular plugin architecture that facilitates extension to alternative solid formulations and additional modeling capabilities. Together, these developments establish a uniquely capable open-source platform for cardiovascular and biomechanics FSI research.

The central contribution of this work is the coupling of two mature, open-source finite element libraries, MFEM and FEBio, each providing complementary capabilities essential for biological FSI applications. MFEM, developed at Lawrence Livermore National Laboratory, provides a scalable, GPU-ready infrastructure designed for exascale computing with support for distributed-memory parallelization as well as hardware accelerators (e.g. GPUs) enabling computing on both traditional and emerging architectures \cite{anderson_mfem_2021,andrej_high-performance_2024}. FEBio, developed at the University of Utah and Columbia University, offers a comprehensive suite of biomechanics modeling features including hyperelastic and viscoelastic constitutive models appropriate for soft tissues, robust contact algorithms essential for leaflet coaptation, and specialized element formulations for nearly-incompressible materials \cite{maas_febio_2012}. This coupling creates a unique synergy: the fluid solver leverages MFEM's parallel computing infrastructure while the immersed solid exploits FEBio's wide range of constitutive modeling as well as advanced computational solid mechanics formulations. Unlike existing immersed FSI frameworks that require users to implement specialized biomechanics features, this approach provides these capabilities out of the box through FEBio's established and validated codebase. While reduced order and simplified structural models decrease computational costs and can be effective for investigations focused on fluid performance, such approaches inadequately capture the tissue mechanics fundamental to understanding disease progression, calcific degeneration, mechanobiological response, and device-tissue interactions \cite{ayoub_heart_2016,el-tallawi_valve_2021,el-tallawi_mitral_2021,marsan_valve_2021,narang_pre-surgical_2021,meador_tricuspid_2020}. The solid modeling capabilities provided via FEBio are essential for applications where tissue-level mechanics in combination with fluid dynamics drive clinical outcomes, including native valve repair optimization, bioprosthetic valve durability prediction, and transcatheter device evaluation \cite{kaiser_simulation-based_2024,martin_comparison_2015,kostyunin_degeneration_2020}.

In this work, we utilize an immersed FSI method based on \cite{black_immersed_2025} and extended to facilitate simpler coupling between FEBio and MFEM. This formulation belongs to the family of fictitous domain/distributed Lagrange multiplier methods, wherein the solid is immersed into a background fluid domain and FSI coupling is enforced as a constraint. For the fluid, we employ a VMS discretization to improve accuracy and robustness on under‑resolved grids, an expected challenge with unfitted meshes in immersed FSI, and to handle the wide range of Reynolds numbers typical of biological flows. Additionally, we include a modification of the stabilization coefficients presented in \cite{kamensky_immersogeometric_2015} to enhance accuracy for large pressure jumps across the fluid-solid interface, a characteristic feature of heart valve FSI dynamics. For the solid, we developed an immersed solid mechanics solver in FEBio with a modular design to facilitate future extensions. The fluid and solid solvers are monolithically coupled resulting in a fully implicit FSI approach. This implicit, monolithic scheme provides robust FSI coupling necessary for strongly coupled fluid-solid interactions commonly observed in biomechanics applications. We utilize the matrix reduction technique from \cite{black_immersed_2025} to reduce the system size to that of a fluid-only problem, thus decreasing the overall computational cost. Furthermore, we iteratively solve this reduced matrix system using block preconditioning techniques originally developed for incompressible flow problems \cite{ryan_t_black_computational_2024,deparis_parallel_2014,liu_nested_2020,benzi_numerical_2005,quarteroni_factorization_2000,elman_taxonomy_2008}.

The immersed FSI method was implemented using FEBio's plugin interface \cite{maas_plugin_2018} and utilizes a modular design that facilitates straightforward extension to alternative solid formulations and additional modeling capabilities. The base class \texttt{MFEMImmersedElasticDomain} (Appendix \ref{ax:secA2}) defines virtual functions for computing element residual vectors and tangent matrices, enabling users to implement specialized element technologies, such as the mean dilatation method for locking-free nearly-incompressible behavior, without modifying core framework components. This design philosophy mirrors FEBio's successful plugin framework, which has enabled community contributions spanning constitutive models, boundary conditions, and solver algorithms \cite{maas_plugin_2018}. The modular structure also facilitates future integration of additional physics, including electromechanical coupling for cardiac simulations and reactive transport for thrombosis modeling. Furthermore, the framework inherits MFEM's state-of-the-art parallel computing capabilities, enabling distributed-memory parallelization via MPI with a clear pathway to GPU acceleration \cite{anderson_mfem_2021}. MFEM has a large feature set, supporting a wide range of applications; discretizations types, including Galerkin, discontinuous Galerkin (DG), mixed methods, isogeometric analysis, hybridization, and discontinuous Petrov-Galerkin approaches; arbitrary high-order finite elements; the full de Rham complex of finite element spaces in 2D and 3D; a wide range of mesh types and support for mesh refinement and optimization; integration with external numerical libraries, such as PETSc \cite{balay_petsctao_2023} and \textit{hypre} \cite{noauthor_hypre_2018}, for additional linear and nonlinear solvers; and massively parallel scalability, HPC efficiency, and GPU acceleration supporting a variety of architectures. We refer the reader to the following citations for more details \cite{anderson_mfem_2021,andrej_high-performance_2024}. This infrastructure positions the framework to benefit from ongoing MFEM development efforts targeting exascale computing, including GPU-accelerated linear solvers and matrix-free operator evaluation as well as scalable automatic differentiation for complex applications \cite{andrej_high-performance_2024,andrej_scalable_2025,usdoe_national_nuclear_security_administration_nnsa_differentiable_2024}. 

The broad capabilities of our immersed FSI solver were demonstrated through a series of numerical test problems. The first example verified solver accuracy via a convergence study using a benchmark problem involving an immersed anisotropic annular solid with an analytical solution. The second example examined an idealized heart valve in both the open and closed configurations to assess performance under the two extreme states characteristic of valve FSI. We then simulated the oscillations of a flexible leaflet in pulsatile cross‑flow, demonstrating the ability of the framework to handle large solid deformations for a problem that would be challenging to model using an ALE-FSI technique and most likely require remeshing to complete the simulation. A free‑falling sphere problem was also included to assess the accuracy of the immersed FSI formulation in three dimensions. Finally, we demonstrated the ability of the proposed framework to perform simulations of three‑dimensional heart valve FSI dynamics.

Several limitations of the current framework should be acknowledged. Contact models should be incorporated into our solver to prevent intersections between immersed bodies rather than relying exclusively on FSI kinematics and the lubrication effect to prevent contact. FSI applications involving immersed thin structures, e.g. heart valve FSI analysis, could benefit from the inclusion of shell formulations to improve computational efficiency. Both contact and shell modeling are features readily available in FEBio and would need to be ported into the present framework, although this process is simplified by the modular design of the immersed solid solver described in Section \ref{sec:plugin_design}. As an immersed (unfitted mesh) approach, the accuracy and efficiency of our computational FSI tool could be improved with adaptive mesh refinement (AMR) techniques to dynamically concentrate resolution near the fluid-solid interface. Accurate resolution of near-wall quantities is important for modeling a variety mechanobiological responses, such as vascular growth and remodeling \cite{humphrey_mechanisms_2008,pfaller_fsge_2024}. Additional performance gains could be realized by utilizing the existing GPU support in MFEM, which was not pursued in this work but is planned for future development. Lastly, the current implementation focuses on Newtonian fluids; extension to non-Newtonian models, while straightforward within the present framework due to the existing support in FEBio, has not been completed.

Future development will focus on several key areas. Further parallelization and optimization of the plugin will be pursued leveraging the existing GPU support in MFEM, which will likely require further method, solver, and algorithmic development. Adaptive mesh refinement capabilities will be implemented to enable efficient resolution of flow features near the fluid-solid interface. Contact models, such as the potential based contact formulation existing in FEBio \cite{kamensky_contact_2018}, will be incorporated into the framework and FSI-specific modifications will be investigated to robustly and accurately model seepage of fluid between the contacting bodies \cite{burman_mechanically_2022,frei_numerical_2025,gerosa_mechanically_2024,ager_consistent_2019,burman_nitsche-based_2020}. Further integration between MFEM and FEBio will expand the range of biological FSI problems that can be addressed, including coupled electromechanical-FSI simulations and multiscale tissue modeling.

\section{Conclusion}\label{sec:conclusion}
This work presents a novel open-source immersed FSI framework that couples the parallel computing capabilities of MFEM with the sophisticated biomechanical modeling features of FEBio, addressing a critical need for computational tools capable of simulating large-deformation biomechanics and cardiovascular problems such as heart valve dynamics. By providing an extensible open-source platform with support for distributed-memory parallelization and a pathway to GPU acceleration, this work enables the broader biomechanics community to perform high-fidelity FSI analyses that simultaneously resolve fluid parameters and tissue-level stress/strain distributions essential for understanding disease progression, optimizing surgical repairs, and evaluating medical device performance. 

The code for the MFEMiFSI plugin will be maintained in the publicly accessible FEBio Plugin Repository and associated Git repository \url{https://github.com/JolleyLab/FEBioMFEMiFSI}. The example problems considered in this work can be found in the FEBio Model Repository at \url{https://repo.febio.org:443/permalink/project/135}. All discussions in this work refer to the state of the plugin at the time of submission. We anticipate improvements to the framework after submission; however, readers can always restore the repository to this state using Git.
\backmatter

\bmhead{Supplementary information}

\bmhead{Author contributions} 
\textbf{RTB} and \textbf{SAM} contributed equally to this work.
\textbf{RTB}: Writing - original draft, Writing – review \& editing, Methodology, Investigation, Formal analysis, Software, Visualization, Conceptualization. 
\textbf{SAM}: Writing - original draft, Writing – review \& editing, Methodology, Investigation, Formal analysis, Software, Visualization, Conceptualization.
\textbf{WW}: Methodology, Investigation, Formal Analysis, Conceptualization, Supervision. 
\textbf{JM}: Investigation, Methodology, Writing – review \& editing
\textbf{TK}: Writing – review \& editing, Supervision, Conceptualization.
\textbf{JAW}: Writing – review \& editing, Supervision, Software, Funding acquisition, Conceptualization.
\textbf{MAJ}: Writing – original draft, Writing – review \& editing, Supervision, Resources, Project administration, Funding acquisition, Conceptualization.

\bmhead{Data availability}
The code for the MFEMiFSI plugin will be maintained at the FEBio Plugin Repository and associated Git repository \url{https://github.com/JolleyLab/FEBioMFEMiFSI}. Example problems from this work are available at the FEBio Model Repository \url{https://repo.febio.org:443/permalink/project/135}. Simulation data will be made available upon request.

\bmhead{Acknowledgements}
This work was funded by The Cora Topolewski Pediatric Valve Center at Children's Hospital of Philadelphia, the Topolewski Endowed Chair in Pediatric Cardiology, a CHOP Cardiac Center Innovation Grant,  NIH R01 HL153166(MAJ), R01 GM083925 (SAM, JAW), T32 HL007915 (RTB) and K25 HL168235 (WW). This work was performed under the auspices of the U.S. Department of Energy by Lawrence Livermore National Laboratory under Contract DE-AC52-07NA27344 (LLNL–JRNL–2014818). High‑performance computing resources provided by the Anvil supercomputer at Purdue University through an NSF ACCESS Discover grant (MCH250067) and by the Children's Hospital of Philadelphia are gratefully acknowledged.

\section*{Declarations}
\bmhead{Conflict of interest} The authors declare no conflict of interest.

\begin{appendices}

\section{Computation of the FSI constraint matrix}\label{ax:secA1}
In this section, we present pseudocode for our algorithm to compute the FSI constraint matrix. The algorithm assumes that each MPI rank has a portion of both the fluid and solid domain.
\begin{algorithm}
\caption{$\mathbf{C}^{\lambda}$ Matrix Assembly}
\begin{algorithmic}[1]
\State \textbf{Given:} Set of solid nodes $\{ x_{A}\}_{A\in\omega_i}$ on processor $i$
\Statex
\State \textbf{Start point search from processor $i$}
\Statex
\State \textbf{Found point on processor $j$}
\For{each point found on processor $j$}
\State Store POINT data = \{parametric position $\boldsymbol{\xi}_A$, element number, and processor number $j$\}
\EndFor
\Statex
\State \textbf{Transfer POINT data on processor $i$ to target processor $j$}

\Statex
\State \textbf{On target processor $j$:}
\For{each received point}
    \State Evaluate fluid shape functions for the containing element at the point
    \If{Fluid DOF is owned by this processor}
        \State Store value in the matrix
    \Else
        \State Store DOF data = \{value, global row, global column, owning processor\}
    \EndIf
\EndFor
\Statex
\State \textbf{Transfer DOF data to owning processors}

\Statex
\State \textbf{On owning processor:}
\For{each received DOF}
    \State Store value in the matrix
\EndFor

\end{algorithmic}
\end{algorithm}

\section{Modular design of the immersed solid implementation}\label{ax:secA2}
In Figure \ref{fig:code_imsolid} below, we provide a snapshot of the \texttt{MFEMImmersedElasticDomain} class showing the virtual functions that allow users to easily extend the current implementation to alternative solid formulations and include additional solid modeling capabilities (e.g. contact). These virtual functions define the computation of several parts of the immersed solid weak form and its linearization at the element level.

\begin{figure}[htbp]
  \centering
\includegraphics[width=\columnwidth]{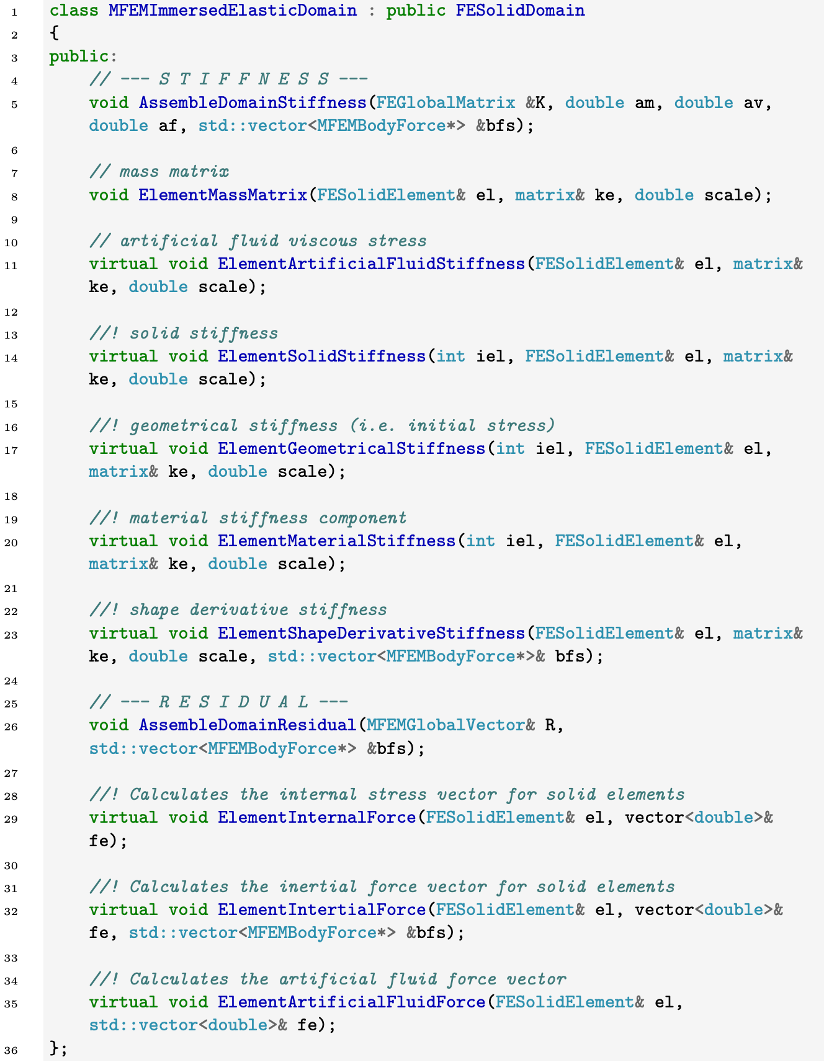}
\caption{Snapshot of the \texttt{MFEMImmersedElasticDomain} class.}
\label{fig:code_imsolid}
\end{figure}

\end{appendices}

\input{main_arXiv.bbl}

\end{document}